\documentclass[runningheads,a4paper]{llncs}

\usepackage{amssymb}
\usepackage{graphicx}

\usepackage[bookmarks,bookmarksopen,bookmarksdepth=3]{hyperref}
\usepackage{microtype}

\usepackage[capitalise,nameinlink]{cleveref}

\usepackage{csquotes} 

\usepackage{caption}
\usepackage{subfigure}

\usepackage[para,online]{threeparttable}

\usepackage[table]{xcolor}

\newcommand{\labeltitle}[1]{\textbf{#1}}

\newcommand{\eg}{e.g.,\ }
\newcommand{\ie}{i.e.,\ }

\crefname{section}{Sect.}{Sect.}
\Crefname{section}{Section}{Sections}
\crefname{figure}{Fig.}{Fig.}
\Crefname{figure}{Figure}{Figures}
\crefname{table}{Tab.}{Tab.}
\Crefname{table}{Table}{Tables}
\crefname{lstlisting}{List.}{List.}
\Crefname{lstlisting}{Listing}{Listings}

\urldef{\mailsa}\path|daniel.ritter@sap.com|

\begin{document}

\mainmatter  

\title{Catalog of Optimization Strategies and Realizations for Composed Integration Patterns}


%
%
\author{Daniel Ritter\inst{1,3},\hspace{-.13cm} Fredrik Nordvall Forsberg\inst{2}\hspace{-.13cm}, Stefanie Rinderle-Ma\inst{3},\hspace{-.13cm} Norman May\inst{1}}
\authorrunning{D.\ Ritter, F.\ Nordvall Forsberg, S.\ Rinderle-Ma and N.\ May}

\institute{SAP SE, Walldorf, Germany, \\
	\email{\{firstname.lastname\}@sap.com}
	\and
	University of Strathclyde, Glasgow, UK,\\
	\email{fredrik.nordvall-forsberg@strath.ac.uk}
	\and
	University of Vienna, Vienna, Austria,\\
	\email{stefanie.rinderle-ma@univie.ac.at}
}

%
%

\maketitle

\begin{abstract}
	The discipline of Enterprise Application Integration (EAI) is the centrepiece of current on-premise, cloud and device integration scenarios.
	However, the building blocks of integration scenarios, \ie essentially a composition of Enterprise Integration Patterns (EIPs), are only informally described, and thus their composition takes place in an informal, ad-hoc manner.
	This leads to several issues including a currently missing optimization of application integration scenarios.
	
	In this work, we collect and briefly explain the usage of process optimizations from the literature for integration scenario processes as catalog.
\end{abstract}

\section{Introduction}

Although Enterprise Application Integration (EAI) \cite{Linthicum:2000:EAI:328930} plays an important role in current IT infrastructures, the development of integration scenarios -- essentially a composition of the Enterprise Integration Patterns (EIP) \cite{hohpe2004enterprise} and extensions \cite{Ritter201736} -- remains a challenging task for the users.
During the modeling of these scenarios, which is currently done with vendor-specific languages and tools, the users have hardly any support or guidance, when it comes to consistency and correction checks of their models, not to mention optimization possibilities \cite{Ritter201736}.

\begin{figure}[bt]
    \centering
    \includegraphics[width=1\columnwidth]{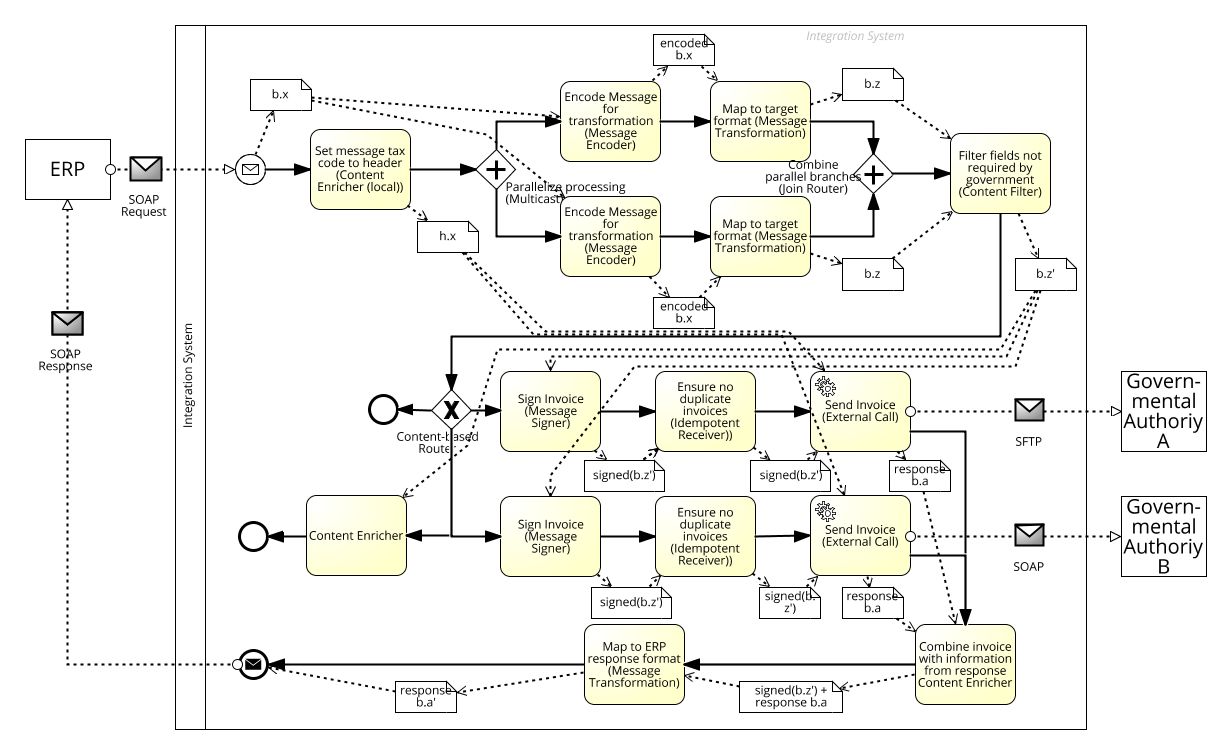}
    \caption{Country-specific invoice processing}
    \label{fig:invoice_italy}
\end{figure}

For instance, \cref{fig:invoice_italy} shows one possible implementation of the \emph{Italy Invoicing Scenario} (country-specific), which allows organizations to communicate with the Italian authorities -- in Business Process Model and Notation (BPMN) with BPMN Message and BPMN Data Object representing the message data flow \cite{DBLP:conf/ecmdafa/Ritter14,DBLP:conf/caise/0001H15,ritter2016exception}.
An organization sends the invoice data, \eg from its ERP system, to the integration process, in which data from the message body $b.x$ to set the required tax data header of the organization required by the governmental authorities using a Content Enricher \cite{hohpe2004enterprise}.
Then the invoice data is mapped to the required target format (Message Transformation \cite{hohpe2004enterprise}) by a mapping program that requires the message in a special encoding (Message Encoder \cite{Ritter201736}).
These operations are parallelized by a Multicast \cite{Ritter201736} and structurally joined by a Join Router \cite{Ritter201736}.
The unnecessary fields of the mapped invoice are removed from the message by a Content Filter \cite{hohpe2004enterprise} resulting to a message content $b.z'$.
The message is routed to one of two receiving governmental authorities, depending on the type of inquiry, by a Content-based Router \cite{hohpe2004enterprise} or discarded (incl. content enricher).
Both governmental authorities require signed messages to verify the authenticity and identity of the sending organization using a Message Signer \cite{Ritter201736}.
Duplicate invoices have to be removed using a (stateful) Idempotent Receiver pattern \cite{hohpe2004enterprise}.
The invoice messages are sent to the authority by an External Call \cite{Ritter201736} pattern and the response is combined with the original message using a content enricher.
The combined data $b.a$ is then mapped to the response format of the ERP sender using a message transformation.

While this example is syntactically and semantically correct, it is not optimal with respect its control and data flow.
Even with an explicit data flow, it is difficult for the user, \eg to recognize the redundant control flows and sibling patterns as well as the potential for pattern parallelization or the benefits of pattern re-ordering for the message throughput.
Furthermore, the modeling complexity and processing latency reduction potential of $\frac{1}{3}$ of the patterns remains hidden to the user (cf. compare with \cref{fig:invoice_simplified_parallel_data_place} in \cref{sec:case_study}).

In this work, we discuss EAI optimization objectives in \cref{sec:objectives} and collect relevant optimization techniques from the literature of related domains of business process, workflow and data integration in an optimization catalog and specify them for EAI processes in \cref{sec:catalog}.
The we discuss the realization of optimization strategies in \cref{sec:realizationold}.
Then we show the applicability to an extended scenario \cref{sec:case_study} and conclude in \cref{sec:discussion}.

\section{Optimization Objectives and Strategies} \label{sec:objectives}

\begin{figure}[bt]
	\centering
	\includegraphics[width=.9\columnwidth]{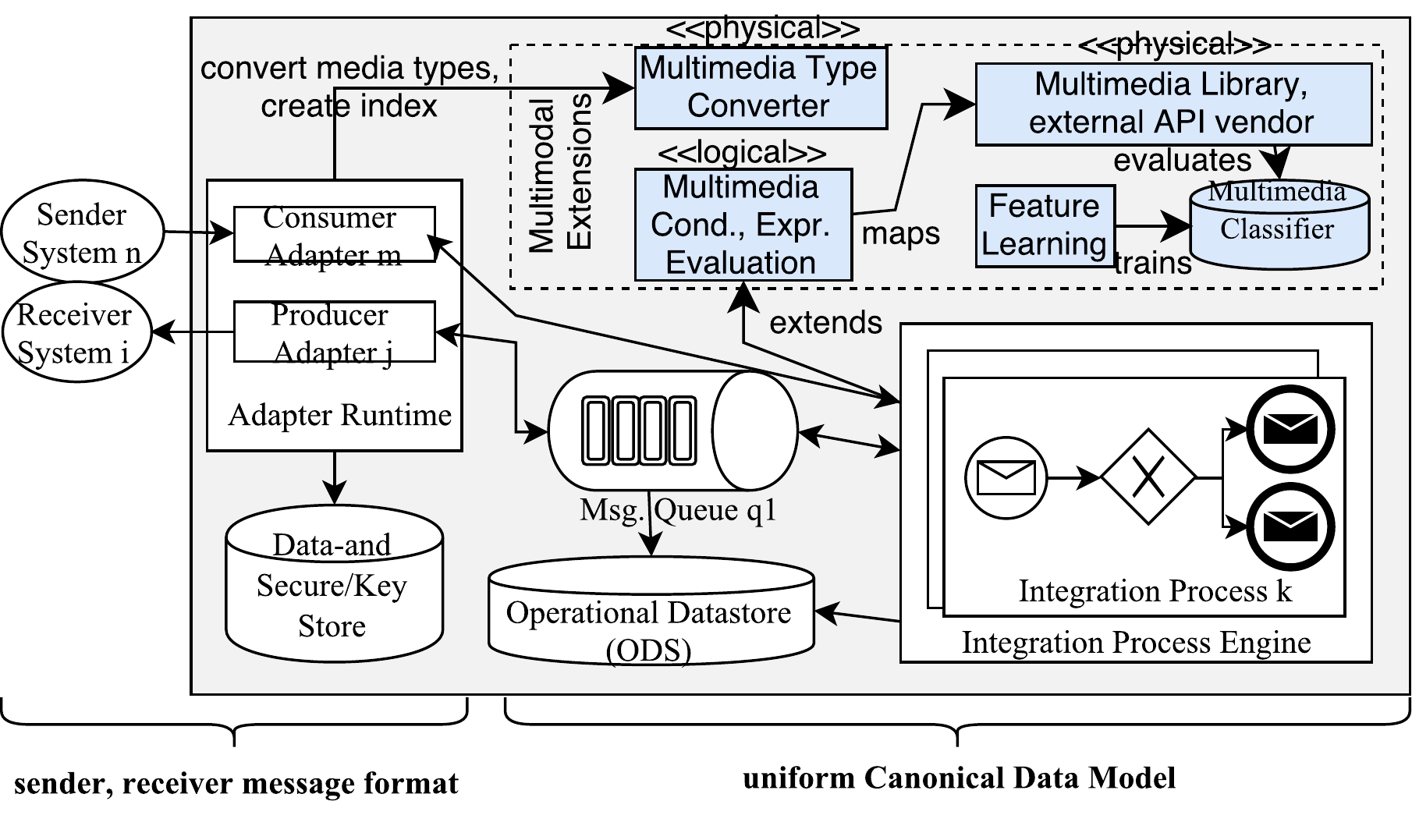}
	\caption{Classical EAI system architecture \cite{DBLP:conf/caise/0001H15} with extensions \cite{edoc2017}}
	\label{fig:eai-system-architecture}
\end{figure}

In this section we collect and discuss EAI optimization objectives in the context of classical EAI \cite{Linthicum:2000:EAI:328930,hohpe2004enterprise} and emerging application integration scenarios \cite{Ritter201736}.
The latter results to new EAI challenges and solutions, which are represented in this work by our studies on \enquote{data-aware} message processing solution spaces: dealing with high velocity and increasing message volume through table-centric processing \cite{DBLP:conf/bncod/Ritter15,DBLP:conf/bncod/Ritter17} and streaming on dataflow (hardware) architectures \cite{DBLP:conf/debs/RitterDMR17}, as well as new message format variety aspects in terms of multimedia integration \cite{edoc2017}.
\Cref{fig:eai-system-architecture} a high-level view on the classical system architecture (based on \cite{DBLP:conf/caise/0001H15}), evolved by new components for multimedia integration (from \cite{edoc2017}).
Subsequently, we define and discuss optimization objectives along the different architectural components.

\subsubsection{Message Throughput and Latency}
The message throughput and the request-response latency are the most important optimizaton objectives of EAI systems.
While the message throughput is the number of messages sent and processed from a sender to a receiver (through the whole system), the latency is the time a sender has to wait until it receives a response from the receiver.
We copnsider message throughput and latency on a message processing pattern or a composition of patterns (\eg including Adapter Runtime, Integration Process Engine).
The work on vectorized integration patterns \cite{DBLP:conf/bncod/Ritter15,DBLP:conf/bncod/Ritter17} illustrates the trade-off of immense message throughput gains, when processing sets of messages in contrast to a reduced overall latency (throughput $\mapsto$ \textbf{Vectorization}).
Furthermore the message throughput can be increased, through processing messages in multiple parallel sub-processes, \eg separate hardware resources \cite{DBLP:conf/debs/RitterDMR17} ($\mapsto$ \textbf{Parallelization}).
The message stream \cite{DBLP:conf/debs/RitterDMR17} and multimedia integration \cite{edoc2017} showed decreasing message throughput for inceasing message sizes.
While message indexing reached its limits for increasing multimedia data \cite{edoc2017}, keeping message sizes smaller helped throughout the experiments ($\mapsto$ \textbf{Data Reduction}).
This could be achieved through process restructurings.
When the structural process changes reduce the number of process elements or complex fork or join patterns, the latency and process complexity could be improved \cite{DBLP:conf/debs/RitterDMR17} ($\mapsto$ \textbf{Process Simplification}).

\subsubsection{Energy Efficiency and Costs}
While the objectives of energy effiiciency and costs are out of scope for this work, some of the named solutions can help to improve both.
The study on application integration on a dataflow hardware architecture \cite{DBLP:conf/debs/RitterDMR17} showed the trade-off between message throughput and energy consumption as well as costs.
Even in modern data centers with virtualized hardware, the resource and energy costs play a crucial role \cite{}.
While the FPGA hardware reached a high throughput with low energy consumption and costs, the software solutions required more energy on one machine and could only reach the same throughput by adding more computing units, and thus increased the costs ($\mapsto$ \textbf{Data flow architecture}).
On the common von Neumann architectures a reduction of message processors can help to improve the energy consumption -- in CPU-bound messaging scenarios -- which also includes the reduction of parallelized processing units ($\mapsto$ \textbf{Process Simplification}).

\subsubsection{Resilience and Robustness}
Since integration systems are central infrastructure components in current IT infrastructures, they are realized as distributed, high-available systems.
However, as shown in \cref{fig:eai-system-architecture}, the dependencies to internal resources (\eg database, message queuing, machine learning) and the growing number of frequently changing message endpoints (\eg mobile and cloud endpoints) \cite{Ritter201736} has to be dealt with during the message processing.
Not only because network communication is a little less reliable than process-to-process communication \cite{hohpe2004enterprise}, smart network usage and reaction to exceptional situations or unavailabilities are crucial ($\mapsto$ \textbf{Reduce Interaction}).

\subsubsection{Context-Awareness and Separation of Duties}
While smarter network usage and resilience improves the stability of integration scenarios, even more distributed processing in context-aware scenarios close to the data producers allows for a short circuit procesing and potentially a reduction of message sizes.
Furthermore -- especially in foreign-hosted, cloud computing environments -- the aspect of secure and private execution of integration processes with user defined functions is crucial \cite{Ritter201736}.
For instance, a process of one cooperation must not be able to interfere with one process of another cooperation.
This requires an integration process fragmentation and placement on heterogeneous runtime systems -- with a focus on reduced interactions ($\mapsto$ \textbf{Pattern Placement}).

\subsubsection{Modeling Complexity}
As seen in \cref{fig:invoice_italy}, even moderately simple integration scenarios can become complex.
One objective could be to build smarter editors that support the users, during modeling, configuration, navigating and understanding an integration scenario.
The understanding and navigating could be approached by simpler integration scenarios ($\mapsto$ \textbf{Process Simplification}) as well as guided modeling ($\mapsto$ \textbf{Smart Guidance}).
In addition, the modeling context, \eg \enquote{query by sketch} support on visual queries for multimedia data integration \cite{edoc2017}, could improve the interaction on the integration process ($\mapsto$ \textbf{Contextualized Modeling}).

\subsubsection{Objectives and Strategies Summary}
Let us summarize the optimization objectives and strategies relevant for EAI in the context of emerging applications.
The objectives are:
\begin{itemize}
    \item Optimization on static design time and dynamic runtime / workload data
    \item Process / Control flow, Data flow and Interaction optimization (not pattern processing efficiency)
    \item Optimization of Message throughput, latency as well as resilience and Separation of Duties (minor focus on energy efficiency and costs)
    \item (Multi-objective optimization)
\end{itemize}

We do not focus on the objectives of pattern solution efficiency, energy efficiency and costs.
Some parts of the reduce modeling complexity objective will be indirectly addressed, due to our focus on integration process related optimizations.
The resulting solution optimization strategies (OSx) that we want to consider are:
\begin{itemize}
    \item\emph{OS1:} Process simplification
    \item\emph{OS2:} Data reduction
    \item\emph{OS3:} Parallelization
    \item\emph{OS4:} Pattern placement
    \item\emph{OS5:} Reduce interacton
\end{itemize}
We do not consider Vectorization, the architecture shift to Data flow architectures, since they partially target the pattern solution efficiency and we already addressed them in the context of EAI in our recent work \cite{DBLP:conf/bncod/Ritter15,DBLP:conf/bncod/Ritter17,DBLP:conf/debs/RitterDMR17}.
Furthermore, we consider smart guidance and contextualized modeling out of scope, due to our focus on integration process optimizations.

Subsequently, we collect optimizations from related domains, classify them according to our objectives and check, whether they help to realize the optimization strategies.






\section{Optimization Catalog} \label{sec:catalog}

In this section we collect relevant optimization techniques for EAI processes from related domains.
Therefore we briefly discuss the selected domains and literature and then list the derived optimization techniques again with brief descriptions.
Furthermore, we discuss the significance of each technique in the context of the optimization objectives from \cref{sec:objectives}.

\subsection{Domain and Literature Selection}

For EAI, the domains of business processes, workflow management and data integration are of particular interest.
Therefore, we conducted a study based on \cite{kitchenham2004procedures} using \texttt{scholar.google.com} with the following keywords: business process optimization, workflow optimization, data integration optimization (\texttt{allintitle}, no patents, accessible, from $2004$ as the EIPs \cite{hohpe2004enterprise}).


\begin{table}[bth]
    \centering
    \footnotesize
    \caption{Optimizations in related domains - horizontal search}
    \label{tab:systemreview}
    \begin{tabular}{lccll}
        \hline
        \parbox[t]{2.5cm}{Keyword} & hits & selected & Selection criteria & Selected Systems \\
        \hline
        \parbox[t]{2.5cm}{Business Process Optimization} & $159$ & $3$ & \parbox[t]{3cm}{data-aware processes} & \parbox[t]{4cm}{survey \cite{vergidis2008business}, optimization patterns \cite{niedermann2011business,niedermann2011deep}} \\
        \parbox[t]{2.5cm}{Workflow Optimization} & $396$ & $6$ & \parbox[t]{3cm}{data-aware processes} & \parbox[t]{4cm}{instance scheduling \cite{agrawal2010scheduling,bittencourt2011hcoc,tirapat2013cost}, scheduling and partitioning for interaction \cite{ahmad2014data}, scheduling and placement \cite{benoit2012throughput}, operator merge \cite{habib2013adapting}} \\
        \parbox[t]{2.5cm}{Data Integration Optimization} & $61$ & $2$ & \parbox[t]{3cm}{data-aware processes optimization, (no schema-matching)} & \parbox[t]{4cm}{instance scheduling, parallelization \cite{zhang2012cost}, ordering, materialization, arguments, algebraic \cite{getta2011static}} \\
        \parbox[t]{2.5cm}{} & & & &\\        
        \parbox[t]{2.5cm}{Added} & n/a & $9$ & expert knowledge & \parbox[t]{4cm}{business process \cite{vrhovnik2007approach}, workflow survey \cite{kougka2014optimization,DBLP:journals/corr/KougkaGS17}, data integration \cite{bohm2008model}, distributed applications \cite{DBLP:phd/de/Bohm2010,DBLP:journals/is/BohmK11}, EAI \cite{DBLP:conf/bncod/Ritter15,DBLP:conf/debs/RitterDMR17,DBLP:conf/bncod/Ritter17}} \\
        \parbox[t]{2.5cm}{Removed} & - & $1$ &  & \parbox[t]{4cm}{classification only \cite{vergidis2008business}} \\
       \parbox[t]{2.5cm}{} &          &               & &  \\

        \parbox[t]{2.5cm}{Overall} & $616$ & $19$ &  &\\
    \end{tabular}
\end{table}

\subsection{Optimization Approach Summaries}

We summarize and classify the found optimization techniques along the objectives of control and data flow as well as design time (static) and runtime or workload (dynamic) optimizations (similar to \cite{bohm2008model}).
Furthermore, we rate these techniques accourding to their relevance for the introduced system imprlementations in the context of emerging application integration.

\subsubsection{Design Time Optimizations}

The design time optimizations denote process improvement techniques that can be identified only on \enquote{static} process models.
Subsequently, we summarize the found static optimization techniques listed in \cref{tab:static_control_data_flow_optimization} and discuss them in the context of the objectives and strategies.

\begin{table}[]
	\centering
    \footnotesize
	\caption{Static Control (SCx) and Data Flow (SDx) Optimizations}
	\label{tab:static_control_data_flow_optimization}
	\begin{tabular}{ll|ccccc|c|l}
		\hline    	
		ID   & \parbox[t]{2.5cm}{Optimization}                     & OS1 & OS2 & OS3 & OS4 & OS5 & other & \parbox[t]{4.5cm}{References} \\
		\hline
        \rowcolor{gray!25}
		SC1 & \parbox[t]{2.5cm}{Redundant Path Removal}          & $\surd$ & - & - & - & - & - &         \parbox[t]{4.5cm}{Remove redundant control flow \cite{bohm2008model}} \\
        \rowcolor{gray!25}
		SC2 & \parbox[t]{2.5cm}{Dead Path Removal}               & $\surd$ & - & - & - & - & - &         \parbox[t]{4.5cm}{Unreachable Sub-graph Elimination \cite{bohm2008model}}\\
        \rowcolor{gray!25}
		SC3 & \parbox[t]{2.5cm}{Sub-process Inlining}              & $\surd$ & - & - & - & - & - &         \parbox[t]{4.5cm}{Local Sub-process Inlining \cite{bohm2008model}}\\
        \rowcolor{gray!25}
        SC4 & \parbox[t]{2.5cm}{Sub-process Generation}            & $\surd$ & - & - & - & $\surd$ & ($\surd$) & \parbox[t]{4.5cm}{(Work-) flow partitioning \cite{ahmad2014data,DBLP:phd/de/Bohm2010,DBLP:journals/is/BohmK11}}\\
		\hline
		SD1 & \parbox[t]{2.5cm}{Resource Squander}                 & - & - & - & - & - & $\surd$ &         \parbox[t]{4.5cm}{Double Variable Assignment \cite{bohm2008model}} \\
		SD2 & \parbox[t]{2.5cm}{Unnecessary Resource Allocation}   & - & - & - & - & - & $\surd$ &         \parbox[t]{4.5cm}{Unnecessary Variable Declaration / Assignment \cite{bohm2008model}; Eliminate-Unused-Variable, SQL simplifications (\eg Eliminate-Redundant-Attributes, Eliminate-Unused-Attributes, Eliminate-Redundant-Predicates) \cite{DBLP:phd/de/Vrhovnik2011}} \\ 
        \rowcolor{gray!25}
		SD3 & \parbox[t]{2.5cm}{Combine Sibling Patterns}          & $\surd$ & - & - & - & - & - &         \parbox[t]{4.5cm}{Two Sibling Translation Operation / Validaton merging \cite{bohm2008model}; Operator Merge \cite{habib2013adapting}} \\
        \rowcolor{gray!25}
		SD4 & \parbox[t]{2.5cm}{Unnecessary conditional fork}      & $\surd$ & - & - & - & ($\surd$) & - & \parbox[t]{4.5cm}{Unnecessary Switch-Path \cite{bohm2008model}; similar to Eliminate-Unused-Partner \cite{DBLP:phd/de/Vrhovnik2011}} \\
		SD5 & \parbox[t]{2.5cm}{Algebraic}                         & - & - & - & - & - & $\surd$ &         \parbox[t]{4.5cm}{Algebraic Optimization / Simplification \cite{bohm2008model,getta2011static,DBLP:phd/de/Bohm2010,DBLP:journals/is/BohmK11}} \\
	\end{tabular}
    \begin{tablenotes}
        \scriptsize
        covers $\surd$, partially covers ($\surd$), does not cover -; relevant optimization techniques \enquote{gray}.
    \end{tablenotes}
\end{table}

The static control (SCx) optimization techniques mostly support the optimization strategy of process simplification (cf. OS1).
Thereby \emph{Redundant Path Removal} and \emph{Dead Path Removal} techniques identify and remove identical, redundant and non-reachable branches in an integration process.
The \emph{Sub-process Inlining} technique helps to simplify modular processes and reduce the number of sub-process calls, however, reduces modularization and fragmentation.
In contrast \emph{Sub-process Generation} allows for the partitioning of a process, which could simplify the main process, helps to place patterns (cf. OS4), while reducing the need for communications (cf. OS5).
  

The techniques from the static data (SDx) optimizations relevant for integration processes \emph{Combine Sibling Patterns} and \emph{Unnecessary conditional fork} again mostly support the optimization strategy of process simplification (cf. OS1).
Thereby (redundant) sibling patterns and unnecessary conditional fork require additional data to decide that the processed elements are the same, the conditional fork statistics indicate no usage, respectively.
The latter can occur in the variant \emph{Eliminate-Unused-Partner}, which even reduces the interaction with message endpoints (cf. OS5).
The found techniques in the areas of \emph{Resource Squander} and \emph{Unnecessary Resource Allocation} target the resource consumption on a \enquote{source code} and not process level.
Hence, they will not be further considered in the context of this work.
Similarly, the \emph{Algebraic} optimizations and simplifications focus on the integration process configuration or the communication between the integration process pipeline and other integration system resources (\eg datastore).

\subsubsection{Runtime Optimizations}
The runtime optimizations denote process improvement techniques that require (continuous) runtime / workload data.
However, in some cases, abstract costs or experimentally collected runtime data is sufficient, \eg latency, message throughput.
Subsequently, we summarize the found static optimization techniques listed in \cref{tab:dynamic_control_data_flow_optimization} and discuss them in the context of the objectives and strategies.

\begin{table}[]
    \centering
    \small
    \caption{Dynamic Control (DCx) and Data Flow (DDx) Optimizations}
    \label{tab:dynamic_control_data_flow_optimization}
    \begin{tabular}{ll|ccccc|c|l}
        \hline    	
        ID   & \parbox[t]{2.5cm}{Optimization}                     & OS1 & OS2 & OS3 & OS4 & OS5 & other & \parbox[t]{4.5cm}{References} \\
        \hline
        DC1 & \parbox[t]{2.5cm}{Message Indexing} 				          & - & ($\surd$) & - & - & - & $\surd$ & \parbox[t]{4.5cm}{2-layer-hash-index \cite{bohm2008model}; ONC \cite{DBLP:conf/bncod/Ritter15}; Feature Selector \cite{edoc2017}; message slices \cite{DBLP:phd/de/Bohm2010,DBLP:journals/is/BohmK11}} \\
        \rowcolor{gray!25}
        DC2 & \parbox[t]{2.5cm}{Distribute Messages} & - & ($\surd$) & ($\surd$) & - & ($\surd$) & - & \parbox[t]{4.5cm}{Heterogeneous load balancing \cite{bohm2008model}}\\
        DC3 & \parbox[t]{2.5cm}{Parallel Process Re-Scheduling} & - & - & - & - & - & $\surd$ & \parbox[t]{4.5cm}{instance scheduling \cite{agrawal2010scheduling,bittencourt2011hcoc,tirapat2013cost}; Rescheduling Start of Parallel Flows \cite{bohm2008model,zhang2012cost}}\\ 
        \rowcolor{gray!25}
        DC4 & \parbox[t]{2.5cm}{Sequence to parallel / merge parallel processes} & - & - & $\surd$ & - & - & - & \parbox[t]{4.5cm}{Rewriting Sequences to Parallel Flows, Merging Parallel Flows \cite{bohm2008model,niedermann2011business,zhang2012cost}; similar to SQL-Parallelization in \cite{DBLP:phd/de/Vrhovnik2011}}\\
        DC5 & \parbox[t]{2.5cm}{Loop to parallel processes} & - & - & $\surd$ & - & - & - & \parbox[t]{4.5cm}{Rewriting Iterations to Parallel Flows \cite{bohm2008model}}\\
        \hline
        DD1 & \parbox[t]{2.5cm}{Reordering / Merging Conditional Paths} & ($\surd$) & - & - & - & - & $\surd$ & \parbox[t]{4.5cm}{Reordering / Merging Switch-Paths \cite{bohm2008model}} \\
        \rowcolor{gray!25}
        DD2 & \parbox[t]{2.5cm}{Pushdown to Endpoints} & - & - & - & $\surd$ & - & - & \parbox[t]{4.5cm}{mentioned in \cite{bohm2008model}; Web-Service-Pushdown to DB \cite{DBLP:phd/de/Vrhovnik2011}} \\
        \rowcolor{gray!25}
        DD3 & \parbox[t]{2.5cm}{Early-Filter} & - & $\surd$ & - & - & - & - & \parbox[t]{4.5cm}{Early Selection \cite{bohm2008model,getta2011static,habib2013adapting,niedermann2011business}; similar to Predicate-Pushdown in \cite{DBLP:phd/de/Vrhovnik2011}} \\
        \rowcolor{gray!25}
        DD4 & \parbox[t]{2.5cm}{Early-Mapping} & - & $\surd$ & - & - & - & - & \parbox[t]{4.5cm}{Early Projection \cite{bohm2008model,getta2011static,habib2013adapting}} \\
        \rowcolor{gray!25}
        DD5 & \parbox[t]{2.5cm}{Early-Aggregation} & - & $\surd$ & - & - & - & - & \parbox[t]{4.5cm}{Early GroupBy \cite{bohm2008model,getta2011static,habib2013adapting}} \\
        \rowcolor{gray!25}
        DD6 & \parbox[t]{2.5cm}{Claim Check} & - & ($\surd$) & - & - & - & - & \parbox[t]{4.5cm}{Materialization Point Insertion / Elimination \cite{bohm2008model,getta2011static}} \\
        DD7 & \parbox[t]{2.5cm}{Algebraic} & - & - & - & - & - & $\surd$ & \parbox[t]{4.5cm}{Orderby Insertion / Removal, Join-Type Selection, Join Enumeration \cite{bohm2008model}; similar to Join-Optimization \cite{DBLP:phd/de/Vrhovnik2011}; message slices \cite{DBLP:phd/de/Bohm2010,DBLP:journals/is/BohmK11}} \\
        DD8 & \parbox[t]{2.5cm}{Process Vectorization} & - & - & - & - & - & $\surd$ & \parbox[t]{4.5cm}{Setoperation-Type Selection, process vectorization \cite{bohm2008model,DBLP:journals/is/BohmHPLW11}; Tuple-to-Set operations like  Insert-Tuple-To-Set, Update-Tuple-To-Set, Delete-Tuple-To-Set \cite{DBLP:phd/de/Vrhovnik2011}; ONC \cite{DBLP:conf/bncod/Ritter15}} \\
        \rowcolor{gray!25}
        DD9 & \parbox[t]{2.5cm}{Pattern Split / Merge} & $\surd$ & - & - & - & - & - & \parbox[t]{4.5cm}{Operator merge \cite{habib2013adapting,bohm2008model,kougka2014optimization,DBLP:journals/corr/KougkaGS17,niedermann2011business}; all Activity-Merging optimizations from \cite{DBLP:phd/de/Vrhovnik2011}: Insert-Update and Update-Insert merging on workload, Select-Merging, Select-Into-Merging; Coarse-Grained-Optimierung \cite{VLDB-2003-KraftSRM}} \\
        DD10 & \parbox[t]{2.5cm}{Pre-computation of Values} & - & - & - & - & - & $\surd$ & \parbox[t]{4.5cm}{\cite{bohm2008model,getta2011static}; Multi-Query-Optimierung techniques \cite{Finkelstein:1982:CEA:582353.582400,Sellis:1986:GQO:16856.16874,Sellis:1988:MO:42201.42203,Park:1988:UCS:645473.653403,Roy:2000:EEA:342009.335419}; Detector Region \cite{edoc2017}} \\
        \rowcolor{gray!25}
        DD11 & \parbox[t]{2.5cm}{Ignore Failing Endpoints} & - & - & - & - & $\surd$ & - & \parbox[t]{4.5cm}{- (Resilience Patterns \cite{Nygard:2007:RDD:1200767,ritter2016exception,Ritter201736})} \\
        \rowcolor{gray!25}
        DD12 & \parbox[t]{2.5cm}{Reduce Requests} & - & ($\surd$) & - & - & $\surd$ & - & \parbox[t]{4.5cm}{Workflow-Database optimization \cite{vrhovnik2007approach}} \\
    \end{tabular}
    \begin{tablenotes}
        \scriptsize
        covers $\surd$, partially covers ($\surd$), does not cover -; relevant optimization techniques \enquote{gray}.
    \end{tablenotes}
\end{table}

The dynamic control (DCx) optimization techniques mostly support process parallelization (cf. OS3) and data reduction (cf. OS2).
The most important one is \emph{Sequence to parallel / merge parallel processes}, which can be considered as technique that elastically scales and reduces parallel patterns or sub-processes.
However, the \emph{Distribute Message} technique can help to reduce interaction with a specific endpoint.
The \emph{Message Indexing} can help to reduce the data, however, rather targets more efficient access data access and the \emph{Parallel Process Re-Scheduling} allows for a different start order of integration processes.
Due to our focus on integration process optimization and our recent work \cite{DBLP:conf/bncod/Ritter15,edoc2017}, these techniques are not further considered.
The same applies to \emph{Loop to parallel processes}, since unfoldable loops are considered a special case in integration processes that would mostly impact patterns like Message Redelivery on Exception~\cite{ritter2016exception}.

The dynamic data (DDx) optimization techniques mostly target data reduction (cf. OS2), \eg \emph{Early-Filter}, \emph{Early-Mapping}, \emph{Early-Aggregation}, and reduced interaction (cf. OS5), \eg \emph{Ignore Failing Endpoints} and \emph{Reduce Requests}.
The data reducing optimizations go along with a process re-ordering, which can even lead to an off-loading to a message endpoint, \eg \emph{Execution Pushdown to Endpoints} (cf. OS-4).
The interactions can be reduced, \eg by ignoring unreachable, non-responsive endpoints or frequently failing endpoints, as well as reducing the number of requests.

\subsection{Optimizations Relevant for Application Integration}

In this section, we assess the practical relevance or the found optimizations for EAI.
We select only those optimizations contributing to strategies OS1--5, which makes the analysis comprehensive only for optimizations on processes (not \enquote{other}).
Then we conduct an impact analysis of the selected optimizations by setting them into context of the objectives as depicted in \cref{tab:relevant_opt_techniques_objectives}.

\begin{table}[]
    \centering
    \small
    \caption{Relevant Optimization Techniques in the Context of the Objectives}
    \label{tab:relevant_opt_techniques_objectives}
    \begin{tabular}{ll|ccccc}
        \hline    	
        ID   & \parbox[t]{4cm}{Optimization}                     & Throughput & Latency & Resilience & Context & Complexity \\
        \hline
        SC1 & \parbox[t]{4cm}{Redundant Path Removal}          & - & $\surd$ & ($\surd$) & - & $\surd$ \\
        SC2 & \parbox[t]{4cm}{Dead Path Removal}               & - & $\surd$ & ($\surd$) & - & $\surd$ \\
        SC3 & \parbox[t]{4cm}{Sub-process Inlining}              & - & $\surd$ & - & - & ($\surd$) \\
        SC4 & \parbox[t]{4cm}{Sub-process Generation}			 & - & - & - & $\surd$ & ($\surd$) \\
        \hline
        SD3 & \parbox[t]{4cm}{Combine Sibling Patterns}          & - & $\surd$ & ($\surd$) & - & $\surd$ \\
        SD4 & \parbox[t]{4cm}{Unnecessary conditional fork}      & $\surd$ & $\surd$ & ($\surd$) & - & $\surd$ \\
        \hline
        \hline
        DC2 & \parbox[t]{4cm}{Distribute Messages} 				      & $\surd$ & - & ($\surd$) & ($\surd$) & - \\
        DC6 & \parbox[t]{4cm}{Sequence to parallel / merge parallel processes} & $\surd$ & ($\surd$) & - & - & - \\
        \hline
        DD1 & \parbox[t]{4cm}{Merging of Switch-Paths} 			      & - & $\surd$ & ($\surd$) & - & $\surd$ \\
        DD2 & \parbox[t]{4cm}{Pushdown to Endpoints} & $\surd$ & ($\surd$) & - & $\surd$ & $\surd$ \\
        DD3 & \parbox[t]{4cm}{Early-Filter}						  & $\surd$ & ($\surd$) & - & ($\surd$) & - \\
        DD4 & \parbox[t]{4cm}{Early-Mapping}                       & $\surd$ & ($\surd$) & - & ($\surd$) & - \\
        DD5 & \parbox[t]{4cm}{Early-Aggregation}                          & $\surd$ & ($\surd$) & - & ($\surd$) & - \\
        DD6 & \parbox[t]{4cm}{Claim Check}        & & & & & \\
        DD9 & \parbox[t]{4cm}{Splitting / Merging Patterns}      & $\surd$ & $\surd$ & - & - & ($\surd$) \\
        DD11 & \parbox[t]{4cm}{Ignore Failing Endpoints} & - & ($\surd$) & $\surd$ & - & - \\
        DD12 & \parbox[t]{4cm}{Reduce Requests} & - & - & $\surd$ & - & - \\
    \end{tabular}
    \begin{tablenotes}
        \scriptsize
        improves $\surd$, possible improvement ($\surd$), does not improve or makes it worse -.
    \end{tablenotes}
\end{table}

In summary, the static control flow optimization techniques act as Process Simplification (cf. OS1), which improves the latency and reduce the modeling complexity.
The sub-process generation technique allows for a more optimal separation of duties.
The static data flow optimization techniques are useful to reduce the latency and modeling complexity.
The dynamic control flow optimization techniques are part of the parallelization strategy (cf. OS3) and vastly impact the message throughput.
Finally, the dynamic data flow optimization techniques are data reducing (cf. OS2), and thus mostly increase the message throughput.
Some of the optimizations might improve the latency and the modeling complexity.
The ignore failing endpoints and reduce requests optimizations improve the resilience of the integration process.



\section{Realization of Optimization Strategies} \label{sec:realizationold}
In this section, we show the realization of the optimizations strategies by discussing the integration patterns that can be used for realizing them.
We also discuss the execution order of the optimization strategies and a decision tree that helps to identify the fitting strategy as practical aspects of the realization.

\subsection{Realization of Optimization Strategies} \label{sec:realization}


In this section we formally define the optimization techniques from the different identified optimization strategies OS1--OS5 in the form of a rule-based graph rewriting system.
We begin by describing the graph rewriting framework, and subsequently apply it to define the optimizations.

\subsubsection{Graph Rewriting}

Graph rewriting provides a visual framework for transforming graphs in a rule-based fashion.
A graph rewriting rule is given by two embeddings of graphs $L \hookleftarrow K \hookrightarrow R$, where $L$ represents the left hand side of the rewrite rule, $R$ the right hand side, and $K$ their intersection (the parts of the graph that should be preserved by the rule).
A rewrite rule can be applied to a graph $G$ after a match of $L$ in $G$ has been given as an embedding $L \hookrightarrow G$; this replaces the match of $L$ in $G$ by $R$.
The application of a rule is potentially non-deterministic: several distinct matches can be possible~\cite{Ehrig:2006:FAG:1121741}.
Visually, we represent a rewrite rule by a left hand side and a right hand side graph colored green and red: green parts are shared and represent $K$, while the red parts are to be deleted in the left hand side, and inserted in the right hand side respectively.
For instance, the following rewrite rule replaces forks with straight edges:
%
%
\begin{center}
\includegraphics[scale=0.5]{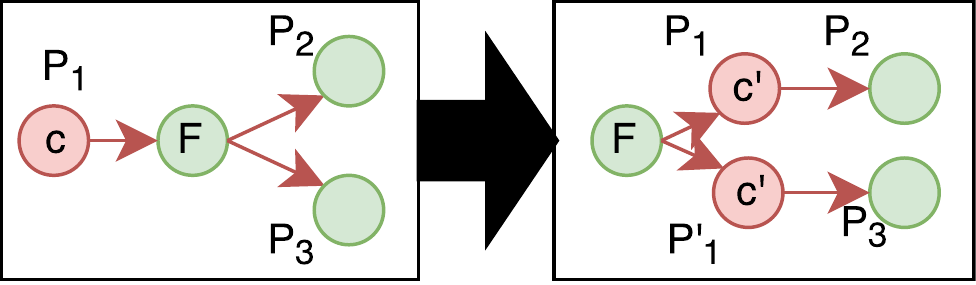}
\end{center}
Formally, the rewritten graph is constructed using a double-pushout (DPO)~\cite{ehrig1973graph} from category theory.
We use DPO rewriting since rule applications are side-effect free (\eg no \enquote{dangling} edges) and local (\ie all graph changes are described by the rules).
We additionally use Habel and Plump's relabelling DPO extension~\cite{habel2002relabelling} to facilitate the relabelling of nodes in partially labelled graphs.
A relabelling is shown in the example, where a property $p$ is rewritten to $p'$.

In addition, we also consider rewrite rules parameterised by graphs, where we draw the parameter graph as a cloud (see e.g.\ \cref{fig:redundant_control_flow_v2} for an example).
A cloud represents any graph, sometimes with some side-conditions that are stated together with the rule.
When looking for a match in a given graph $G$, it is of course sufficient to instantiate clouds with subgraphs of $G$ --- this way, we can reduce the infinite number of rules that a parameterised rewrite rule represents to a finite number.
Parameterised rewrite rules can formally be represented using substitution of hypergraphs~\cite{plump1994hypergraph} or by !-boxes in open graphs~\cite{bangBoxes}.
Since we will describe optimisation strategies as graph rewrite rules, we can be flexible with when and in what order we apply the strategies.
We apply the rules repeatedly until a fixed point is reach, \ie when no further changes are possible.
Methodologically, the rules are specified in a pattern-like formalsim with pre-conditions, change primitives, post-conditions and an optimization effect.
The pre- and post conditions are part of the re-writing rule representation.

\subsubsection{OS-1: Process Simplification}

We first consider the process simplification optimization strategies from \cref{sec:objectives} that mainly strive to reduce the model complexity and latency.


\paragraph{Redundant sub-process} This optimisation removes redundant copies of the same sub-process within a process.

\begin{figure}[bt]
  \subfigure[Redundant sub-process]{\label{fig:redundant_control_flow_v2}\includegraphics[width=0.49\columnwidth]{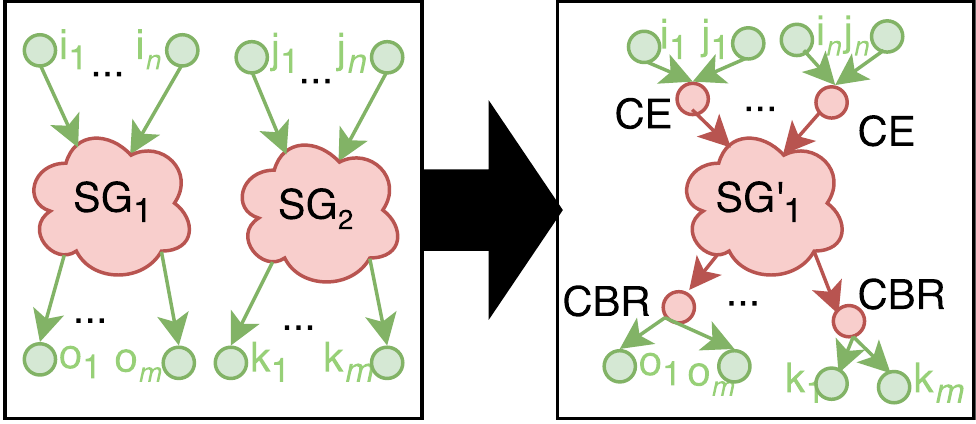}}
  \hfill
  \subfigure[Combine sibling patterns]{\label{fig:combine_sibling_nodes_v2_b}\includegraphics[width=0.49\columnwidth]{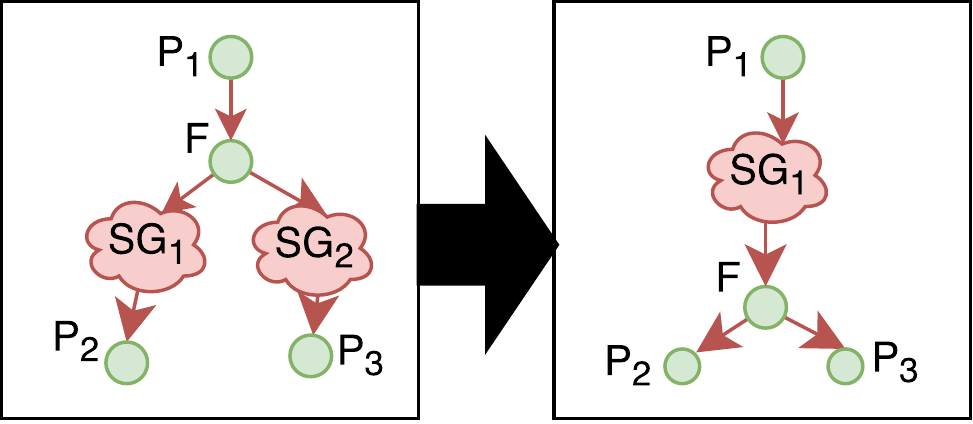}} 
  \caption{Rules for redundant sub-process and combine sibling patterns}
  \label{fig:redundant}
\end{figure}

\labeltitle{Change primitives:} The rewriting is given by the rule in \cref{fig:redundant_control_flow_v2}, where $SG1$ and $SG2$ are isomorphic pattern graphs with in-degree $n$ and out-degree $m$.
In the right hand side of the rule, the $CE$ nodes add the context of the predecessor node to the message in the form of a content enricher pattern, and the $CBR$ nodes are content-based routers that route the message to the correct recipient based on the context introduced by $CE$.
The graph $SG'_1$ is the same as $SG_1$, but with the context introduced by $CE$ copied along everywhere.

\labeltitle{Effect:} The optimization is beneficial for model complexity when the isomorphic subgraphs contain more than $n + m$ nodes, where $n$ is the in-degree and $m$ the out-degree of the isomorphic subgraphs.
The latency reduction is by the factor of subgraphs minus the latency introduced by the $n$ extra nodes $CE$ and $m$ extra nodes $CBR$.

\paragraph{Combine sibling patterns} Sibling patterns have the same parent node in the pattern graph (\eg they follow a non-conditional forking pattern) with implied channel cardinality of $1$:$1$.


\labeltitle{Change primitives:} The rule is given in \cref{fig:combine_sibling_nodes_v2_b}, where $SG_1$ and $SG_2$ are isomorphic pattern graphs that are side-effect free, and $F$ is an unconditional fork.

\labeltitle{Effect:} The model complexity and latency are reduced by the model complexity and latency of $SG_2$.



\paragraph{Unnecessary Fork Paths} A fork path is unnecessary, if the patterns on the path are side-effect free without a transitive connection to a message endpoint, or additionally read-only, when transitively connected to an endpoint.

\begin{figure}[bt]
  \subfigure[Transitive Endpoint]{\label{fig:unnecessary_switch_paths_b}\includegraphics[width=0.49\columnwidth]{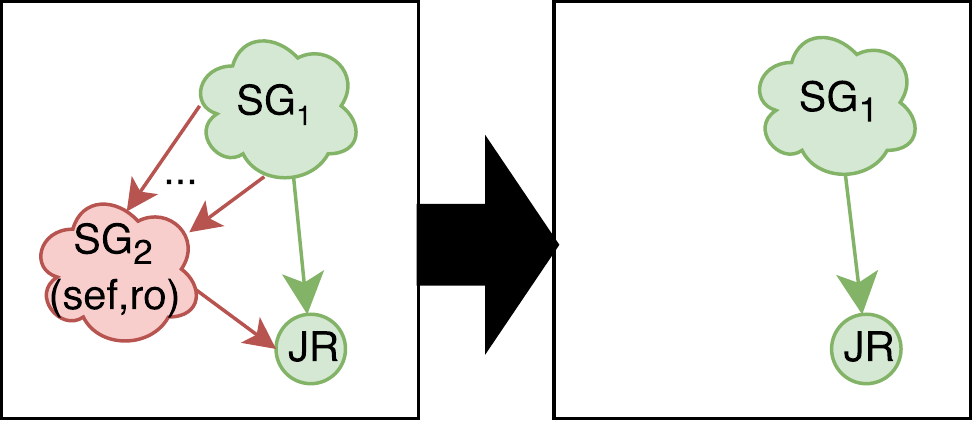}}
  \hfill
  \subfigure[No Endpoint]{\label{fig:unnecessary_switch_paths_c}\includegraphics[width=0.49\columnwidth]{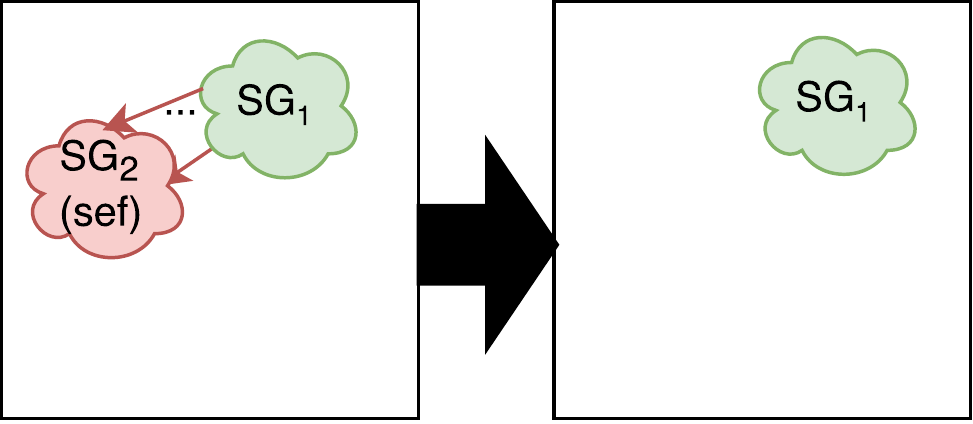}}
  \caption{Rules for unnecessary fork paths}
  \label{fig:unnecessary_switch}
\end{figure}


\labeltitle{Change primitives:} The rules are given in \cref{fig:unnecessary_switch}, where 
$SG_2$ denotes a side-effect free 
and read-only 
subgraph, and $JR$ is a join router.

\labeltitle{Effect:} The model complexity and latency are reduced by the number and costs of the patterns in the subgraph.



\subsubsection{OS-2: Data Reduction, OS-4: Pattern Placement} \label{sub:data_reduction}

Now, we consider data reduction optimization strategies, which mainly target improvements of the message throughput (incl. reducing element cardinalities).
These optimizations require that pattern input and output contracts are regularly updated with snapshots of element data sets $EL_{iC}$ and $EL_{oC}$ from live runtime systems, \eg from experimental measurements through benchmarks~\cite{ritter2016benchmarking}.

All of the data reduction optimizations discussed in this section can also be applied as OS-4 pattern placement strategies (\enquote{Pushdown to Endpoint}), by extending the placement to the message endpoints. 
Due to our focus on integration processes, we will not further elaborate on this here.

\paragraph{Early-Filter} A filter pattern can be moved to or inserted prior to some of its successors to reduce the data to be processed.
The following types of filters have to be differentiated:
\begin{itemize}
  \item A \emph{message filter} removes messages with invalid or incomplete content.
    It can be used 
    to prevent exceptional situations, and thus improves 
    stability. 
  \item A \emph{content filter} removes elements from messages. It can be used to reduce the amount of data passed to subsequent patterns.
\end{itemize}
%

\begin{figure}[bt]
  \subfigure[Early Filter]{\label{fig:early_filter_v2_insert}\includegraphics[width=0.49\columnwidth]{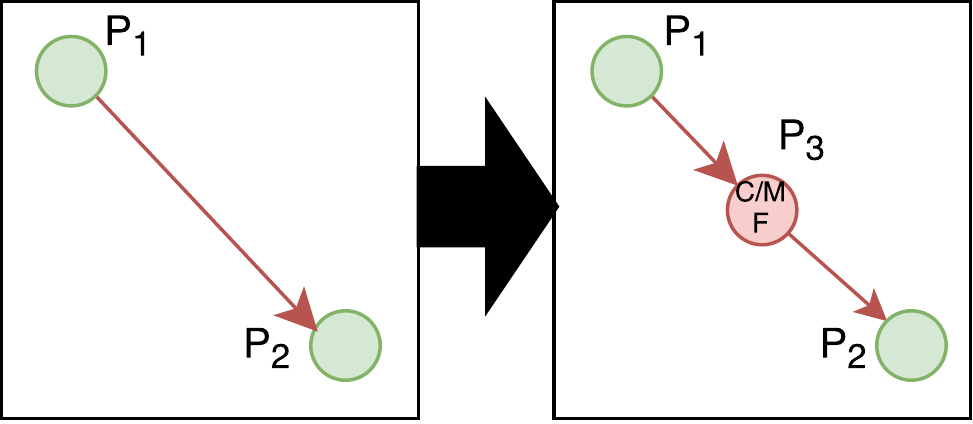}}
  \hfill
  \subfigure[Early Mapping]{\label{fig:early_mapping_v2}\includegraphics[width=0.49\columnwidth]{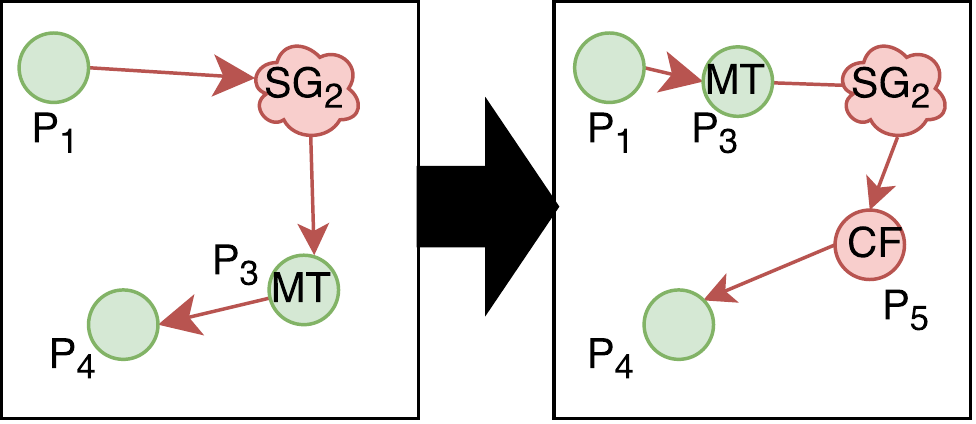}}
  \caption{Rules for early-filter and early-mapping.}
  \label{fig:early_filter}
\end{figure}

\labeltitle{Change primitives:} The rule is given in \cref{fig:early_filter_v2_insert}, where $C/M F$ is either a content or message filter matching the output contracts of $SG_1$ and the input contracts of $P_1$, and filtering out the data not used by $P_1$.

\labeltitle{Effect:} The message throughput increases by the ratio of the number of reduced data elements that are processed per second, if not limited by the throughput of the additional pattern.

\paragraph{Early-Mapping} A mapping that reduces the number of elements in a message can increase the message throughput.


\labeltitle{Change primitives:} The rule is given in \cref{fig:early_mapping_v2}, where $P_1$ is an element reducing message mapping compatible with both $SG_2$, $SG_3$, and $SG_1$, $SG_2$, and where no pattern in $SG_3$ modifies the elements mentioned in the output contract of $P_1$.
Furthermore $P_2$ is a content filter, which ensures that the input contracts of the subsequent patterns in $SG_3$ are satisfied.


\labeltitle{Effect:} The message throughput for the subgraph subsequent to the mapping increases by the ratio of the number of unnecessary data elements that have to be processed.

\paragraph{Early-Aggregation} A micro-batch processing region is a subgraph which contains patterns that are able to process multiple messages combined to a multi-message~\cite{DBLP:conf/bncod/Ritter17} or one message with multiple segments with an increased message throughput.
The optimal number of aggregated messages is determined by the highest batch-size for the throughput ratio of the pattern with the lowest throughput, if latency is not considered.

\begin{figure}[bt]
  \subfigure[Early-Aggregation]{\label{fig:early_aggregate}\includegraphics[width=0.49\columnwidth]{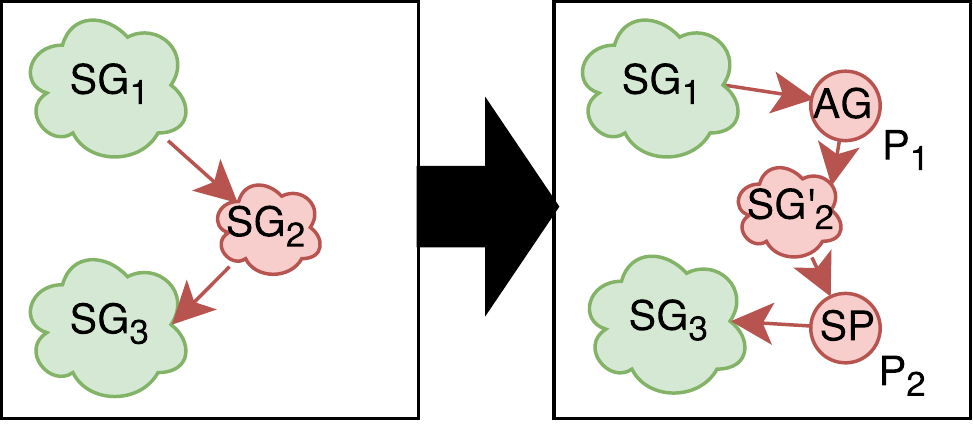}}
  \hfill
  \subfigure[Early-Claim Check]{\label{fig:early_claim_v2}\includegraphics[width=0.49\columnwidth]{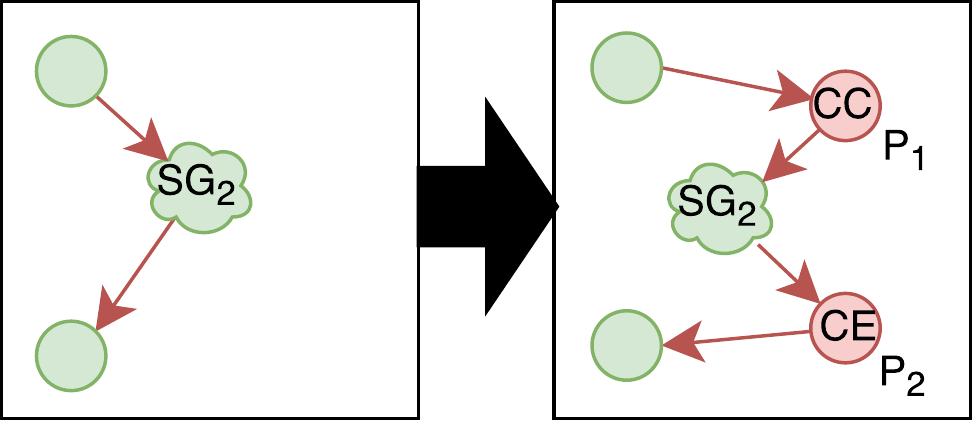}}
  \caption{Rules for early-aggregation and early-claim check}
  \label{fig:early_aggregate_claim}
\end{figure}


\labeltitle{Change primitives:} The rule is given in \cref{fig:early_aggregate}, where $SG_2$ is a micro-batch processing region, $P_1$ an aggregator, $P_2$ a splitter which separates the batch entries to distinct messages to reverse the aggregation, and $SG'_2$ finally is $SG_2$ modified to process micro-batched messages.

\labeltitle{Effect:} The message throughput is the minimal pattern throughput of all patterns in the micro-batch processing region.
If the region is followed by patterns with less throughput, only the overall latency might be improved.

\paragraph{Early-Claim Check}
If a subgraph does not contain a pattern with message access, the message payload can be stored intermediately persistently or transiently (depending on the quality of service level) and not moved through the subgraph.
For instance, this applies to subgraphs consisting of data independent control-flow logic only, or those that operate entirely on the message header (\eg header routing).


\labeltitle{Change primitives:} The rule is given in \cref{fig:early_claim_v2}, where $SG_2$ is a message access-free subgraph, $P_1$ a claim check that stores the message payload and adds a claim to the message properties (and possibly routing information to the message header), and $P_2$  a content enricher that adds the original payload to the message.

\labeltitle{Effect:} The main memory consumption and CPU load decreases, which could increase the message throughput of $SG_2$, if the claim check and content enricher pattern throughput is greater than or equal to the improved throughput of each of the patterns in the subgraph.

\paragraph{Early-Splitter} Messages with many segments can be reduced to several messages with fewer segments, reducing complexity.
Given scenario workload statistics $ws_k$ for scenarios $k$, and an assignment $M(p, ts)$ of number of segments for each process $p$ and throughput statistics $ts$, a segment bottleneck subsequence consists of a set of adjacent patterns $\{p_1, \ldots, p_m\}$ such that their $ws_k$ is signinficantly lower than of their preceding and succeeding patterns.
The patterns in the sequence must have a similar optimal segment size.
Algorithmically, bottlenecks could be found using max flow-min cut techniques.

\begin{figure}[bt]
  \subfigure[Early Split]{\label{fig:early_split_v2}\includegraphics[width=0.49\columnwidth]{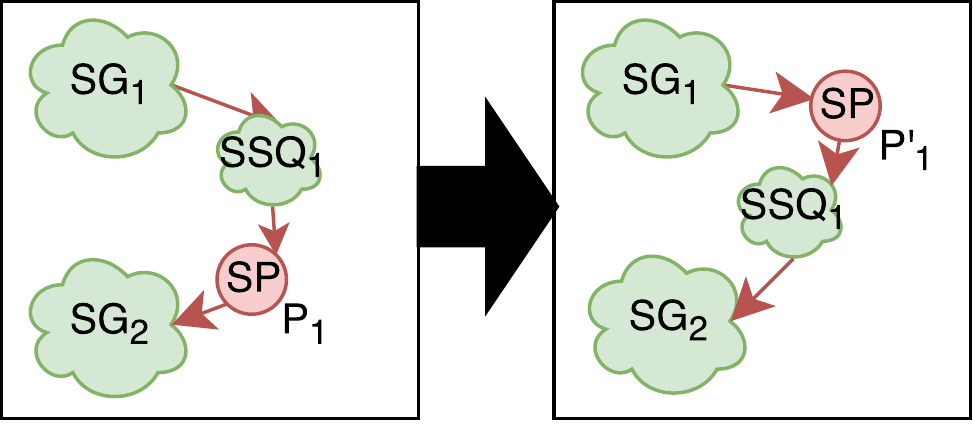}}
  \hfill
  \subfigure[Early Split (inserted)]{\label{fig:early_split_v2_insert}\includegraphics[width=0.49\columnwidth]{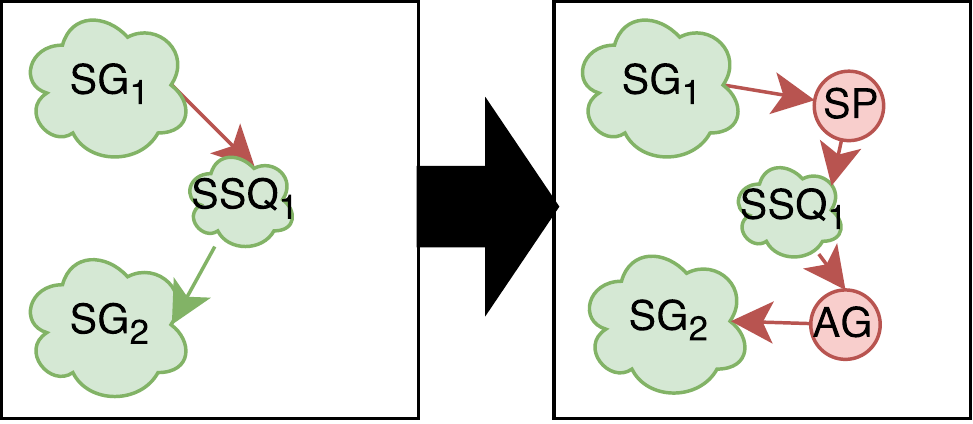}}
    \caption{Rules for early split.}
    \label{fig:early_split}
\end{figure}

\labeltitle{Change primitives:} The rule is given in \cref{fig:early_split}, where $SSQ_1$ is a segment bottleneck sub-sequence.
If $SSQ_1$ already has an adjacent splitter, \cref{fig:early_split_v2} applies, otherwise \cref{fig:early_split_v2_insert}.
In the latter case, $SP$ is a splitter and $P_2$ is  an aggregator that re-builds the required segments for the successor in $SG_2$.
For an already existing splitter $P_1$ in \cref{fig:early_split_v2}, the split condition has to be adjusted to the elements required by the input contract of the subsequent pattern in $SSQ_1$.
In both cases we assume that the patterns in $SSQ_1$ deal with single- and multi-segment messages; otherwise all patterns have to be adjusted as well.


\labeltitle{Effect:} The message throughput increases by the ratio of increased throughput on less message segments minus, if the throughput of the moved or added splitter (and aggregator) $\geq$ message throughput of each of the patterns in the segment bottleneck sub-sequence after the segment reduction.

\subsubsection{OS-3: Parallelization}

Parallelization optimization strategies increase message throughput.
Again, these optimizations are not completely decidable on static integration scenarios according to \cref{sec:objectives} but require experimentally measured message throughput statistics $MT_{c}$ (\eg benchmarks~\cite{cs4681}).

\paragraph{Sequence to parallel}
A bottleneck sub-sequence with channel cardinality 1:1 can also be handled by parallelizing it.
The parallelization factor is the average message throughput of the predecessor and successor of the sequence divided by two, which denotes the improvement potential of the bottleneck sub-sequence.
The goal is to not overachieve the mean of predecessor and successor message throughput with the improvement to avoid iterative re-optimization.

\begin{figure}[bt]
  \subfigure[Sequence to parallel]{\label{fig:sequence_to_parallel_v2}\includegraphics[width=0.49\columnwidth]{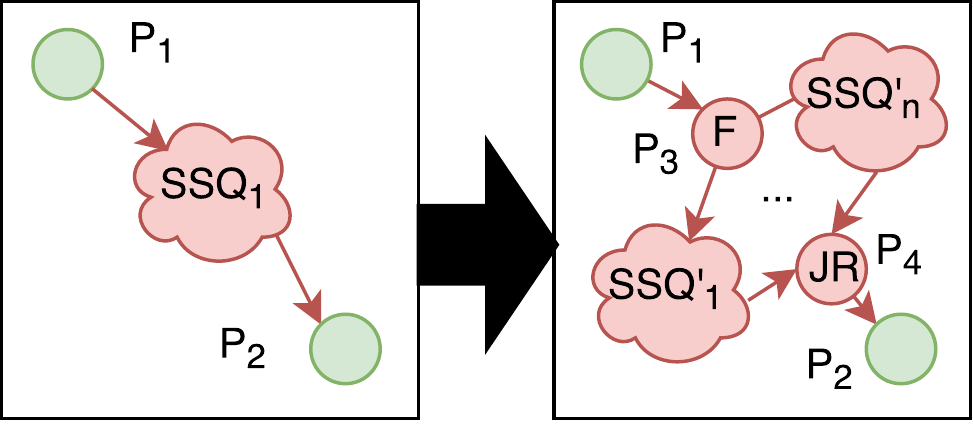}}
  \hfill
  \subfigure[Merge parallel]{\label{fig:merge_parallel_v2}\includegraphics[width=0.49\columnwidth]{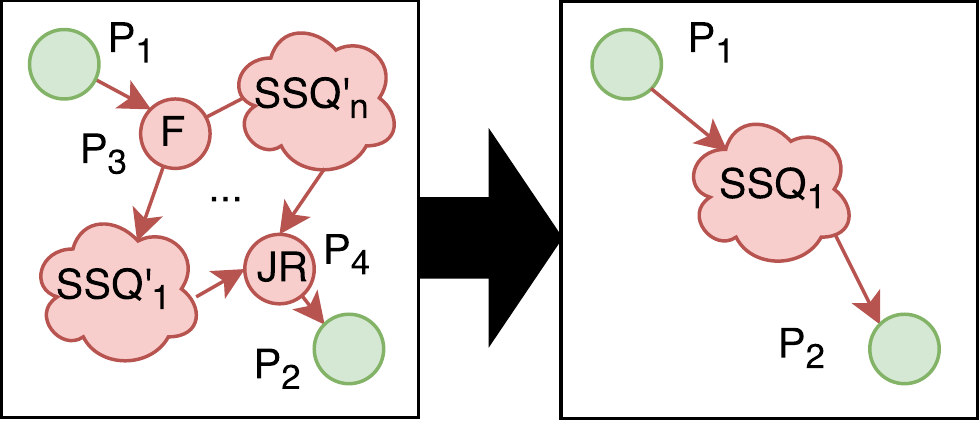}}
	\caption{Rules for sequence to parallel and merge parallel.}
	\label{fig:parallel}
\end{figure}


\labeltitle{Change primitives:} The rule is given in \cref{fig:sequence_to_parallel_v2}, where $SSQ_1$ is a bottleneck sub-sequence, $P_2$ a fork node, $P_3$ a join router, and each $SSQ'_k$ is a copy of $SSQ_1$, for $1 \leq k \leq n$.
The parallelization factor $n$ is a parameter of the rule.


\labeltitle{Effect:} The message throughput improvement rate depends on the parallelization factor $n$, and the message throughput of the balancing fork and join router on the runtime.
For a measured throughput $t$ of the identified bottleneck sub-sequences, the message throughput can be improved to $n \times t \leq$ average sum of predecessor and successor throughput, while limited by upper boundary of balancing fork or join router.

\paragraph{Merge parallel} The balancing fork and join router realizations can limit the throughput in some runtime systems, so that a parallelization decreases the throughput.
This is called a limiting parallelization, and is defined as when a fork or a join has smaller throughput than a pattern in the following sub-sequence.


\labeltitle{Change primitives:} The rule is given in \cref{fig:merge_parallel_v2}, where $P_3$ and $P_4$ limit the message throughput of each of the $n$ sub-sequence copies $SSQ'_1$, \ldots, $SSQ'_n$ of $SSQ_1$.


\labeltitle{Effect:} The model complexity is reduced $(n-1)k - 2$, where each $SSQ'_i$ contains $k$ nodes.
The message throughput might improve, since the transformation lifts the limiting upper boundary of a badly performing balancing fork or join router implementations to the lowest pattern throughput in the bottleneck sub-sequence.

\subsubsection{Heterogeneous Parallelization}
A heterogeneous parallelization consists of parallel sub-sequences that are not isomorphic.
In general, two subsequent patterns $P_i$ and $P_j$ can be parallelized, if the predecessor pattern of $P_i$ fulfills the input contract of $P_j$, $P_i$ behaves read-only with respect to the data element set of $P_j$, and the combined outbound contract of $P_i$ and $P_j$ fulfill the input contract of the successor pattern of $P_j$.

\begin{figure}[bt]
  \subfigure[Heterogeneous sequence to parallel]{\label{fig:heterogeneous_sequence_to_parallel_v2}\includegraphics[width=0.49\columnwidth]{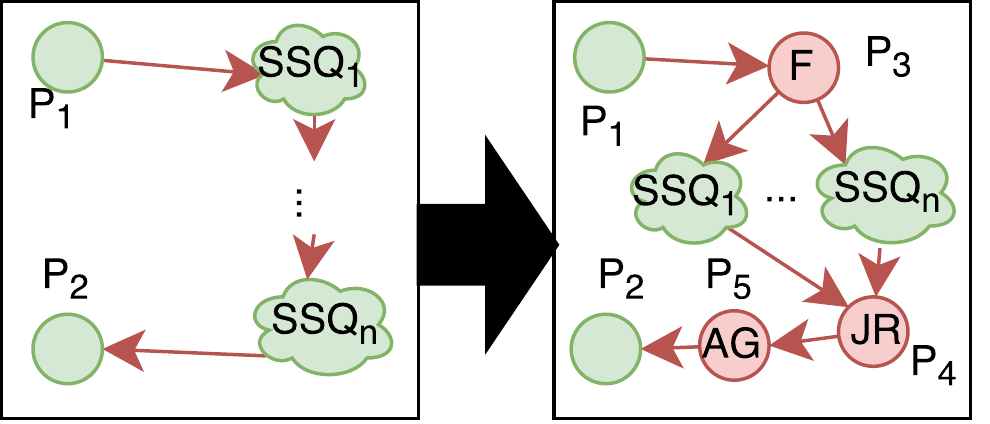}}
  \hfill
  \subfigure[Merge heterogeneous parallel]{\label{fig:heterogeneous_merge_parallel_v2}\includegraphics[width=0.49\columnwidth]{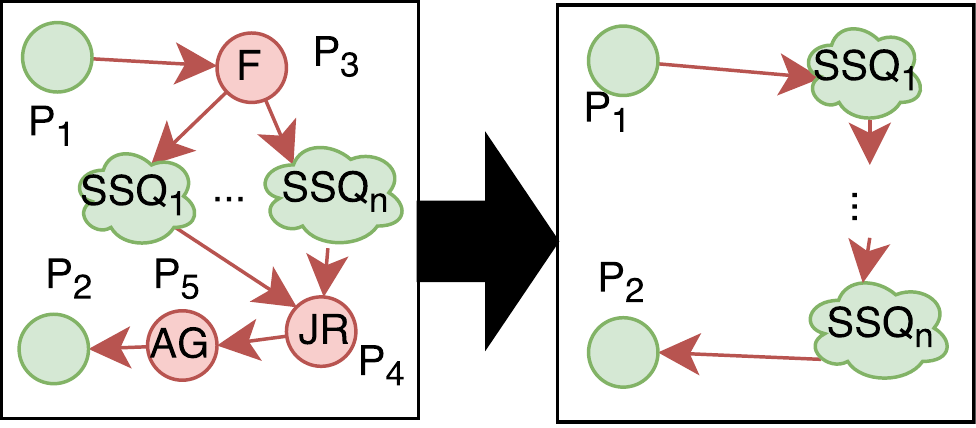}}
    \caption{Rules for heterogeneous parallelization and merging heterogeneous parallel.}
    \label{fig:hetero_parallel}
\end{figure}

\labeltitle{Change primitives:} The rule is given in \cref{fig:heterogeneous_sequence_to_parallel_v2}, where the sequential sub-sequence parts $SSQ_1$, .., $SSQ_n$ can be parallelized, $P_3$ is a parallel fork, $P_4$ is a join router, and $P_5$ is an an aggregator that waits for messages from all sub-sequence part branches before emitting a combined message that fulfills the input contract of $P_2$.
Similarly a rule for a heterogeneous merge is given in \cref{fig:heterogeneous_merge_parallel_v2} (see \emph{Merge parallel} above).

\labeltitle{Effect:}
Synchronization latency can be improved, but the model complexity increases by 3.
The latency improves from the sum of the sequential pattern latencies to the maximal latency of all sub-sequence parts plus the fork, join, and aggregator latencies.







 
\subsubsection{OS-5: Reduce Interaction}

Optimization strategies that reduce interactions target a more resilient behavior of an integration process.

\paragraph{Ignore Failing Endpoints} When endpoints fail, different exceptional situations have to be handled on the caller side.
In addition, this can come with long timeouts, which can block the caller and increase latency.
Knowing that an endpoint is unreliable can speed up processing, by immediately falling back to alternative.

\begin{figure}[bt]
  \subfigure[Ignore Failing Endpoint]{\label{fig:ignore_failing_endpoints_v2}\includegraphics[width=0.49\columnwidth]{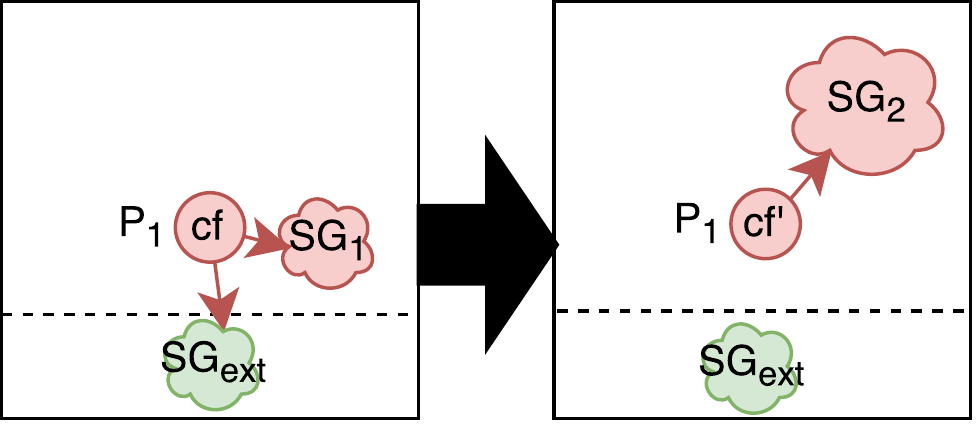}}
  \hfill
  \subfigure[Try Failing Endpoint]{\label{fig:ignore_failing_endpoints_v2_b}\includegraphics[width=0.49\columnwidth]{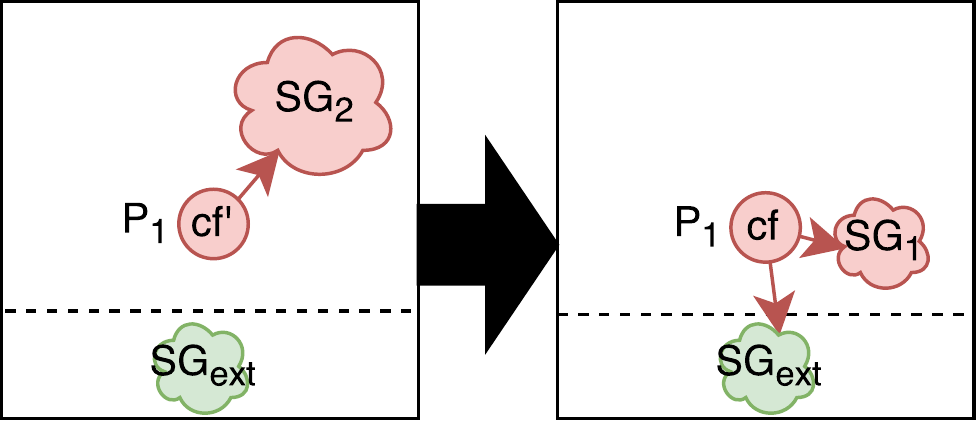}}
    \caption{Rules for ignore failing endpoints.}
    \label{fig:ignore_failing}
\end{figure}


\labeltitle{Change primitives:} The rule is given in \cref{fig:ignore_failing_endpoints_v2}, where $SG_{ext}$ is a failing endpoint, $SG_1$ and $SG_2$ subgraphs, and $P_1$ is a service call or message send pattern  with configuration $cf$.
This specifies the collected number of subsequently failed delivery attempts to the endpoint or a configurable time interval.
If one of these thresholds is reached, the process stops calling $SG_{ext}$ and does not continue with the usual processing in $SG_1$, however, invokes an alternative processing or exception handling in $SG_2$.




\labeltitle{Effect:} Besides and improved latency (\ie average time to response from endpoint in case of failure), the integration process behaves more stable due to immediate alternative processing.
To not exclude the remote endpoint forever, the rule in \cref{fig:ignore_failing_endpoints_v2_b} is scheduled for execution after a period of time to try whether the endpoint is still failing.
If not, the configuration is updated to $cf'$ to avoid the execution of \cref{fig:ignore_failing_endpoints_v2}.
The retry time is adjusted depending on experienced values (\eg endpoint is down every two hours for ten minutes).

\paragraph{Reduce Requests} A \emph{message limited} endpoint, \ie an endpoint that is not able to handle a high rate of requests, can get unresponsive or fail.
To avoid this, the caller can notice this (\eg by TCP back-pressure) and react by reducing the number or frequency of requests.
This can be done be employing a throttling or even sampling pattern~\cite{ritter2016exception}, which removes messages.
An aggregator can also help to combine messages to multi-messages~\cite{DBLP:conf/bncod/Ritter17}.

\begin{figure}[bt]
  \subfigure[Reduce Requests]{\label{fig:reduce_requests_v2}\includegraphics[width=0.49\columnwidth]{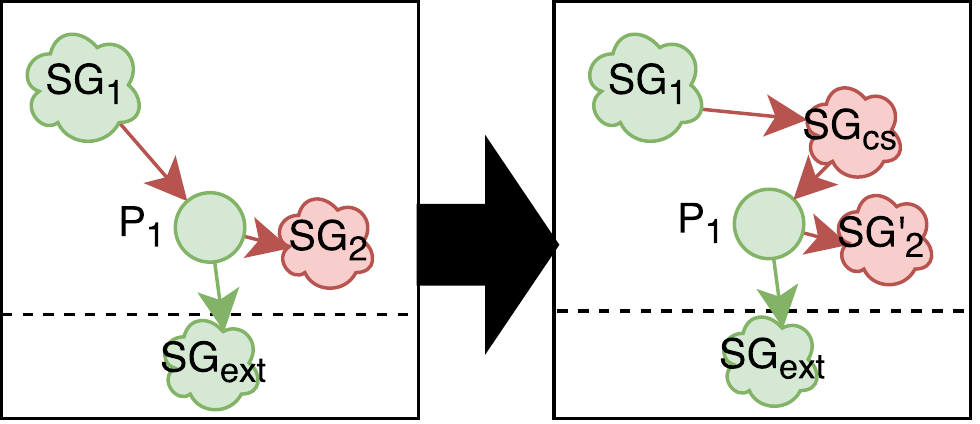}}
  \caption{Rules for reduce requests.}
  \label{fig:reduce}
\end{figure}


\labeltitle{Change primitives:} The rewriting is given by the rule in \cref{fig:reduce_requests_v2}, where $P_1$ is a service call or message send pattern, $SG_{ext}$  a message limited external endpoint, $SG_2$ a subgraph with $SG'_2$ a re-configured copy of $SG_2$ (\eg for vectorized message processing~\cite{DBLP:conf/bncod/Ritter17}), and $SG_{cs}$ a subgraph that reduce the pace, or number of messages sent.


\labeltitle{Effect:} Latency and message throughput might improve, but this optimization mainly targets stability of communication.
This is improved by configuring the caller to a message rate or number of requests that the receiver can handle.



\subsection{Realization by Patterns}
One way of realizing the optimization strategies is to use the integration patterns themselves.
The patterns denote general categories of solutions for message processing, and thus cover a large solution space for the realization of optimizations.
Subsequently, we assign and discuss the applicability of integration patterns to realize optimization strategies denoted in \cref{sec:objectives}.

\begin{table}[]
	\centering
	\caption{Optimization Strategies realized by Integration Patterns}
	\label{tab:optimization_strategy_realization}
	\begin{tabular}{ll}
		Strategy ID & Realization Categories \\
		\hline
		OS-1 & - (structural criteria only) \\
		OS-2 & \parbox[t]{10cm}{Permanent message removal (cf. Early-Filter messages, Early-Mapping) / temporal message removal (cf. Claim Check), message combination (cf. Early-Aggregation), message content filtering (cf. Early-Filter data) / size reduction (cf. Compression)}\\
		OS-3 & \parbox[t]{10cm}{- (supporting Sequence to Parallel, Merge Parallel through non-conditional forking and joining patterns; Heterogeneous Load Balancing through forking and format handling)} \\
		OS-4 & \parbox[t]{10cm}{cf. OS-2 categories, Early-Quality of Service and Early-Operation (cf. pushdown), Late-Quality of Service and Right-Place (cf. push-back)} \\
		OS-5 & \parbox[t]{10cm}{resilience, tolerance (cf. Ignore Failing Endpoints), filtering and flow control (cf. Reduce Requests), prevention (cf. Distribute Messages)} \\
	\end{tabular}
\end{table}


The basic characteristics of pattern compositions (\ie soundness and reachability) as well as process simplifications (cf. OS-1) denote structural criteria on the pattern graph, and thus do not require patterns for their realization.
There are 16 out of 153 patterns with potential of data reduction (cf. OS-2) can be categorized along message and message content reducing patterns.
The message reducing patterns, reduce the amount of data through their potential to remove messages permanently (\eg Content-Based Router, Message Filter, Selective Consumer, Idempotent Receiver, Channel Purger from \cite{hohpe2004enterprise}, Message Cancellation, Message Expiration, Validate Message, Message Sampler from \cite{Ritter201736}) or temporarily (\eg Claim Check \cite{hohpe2004enterprise}), combine messages (\eg Aggregator, Composed Msg. Processor from \cite{hohpe2004enterprise}), or reduce their message content by filtering (\eg Splitter, Content Filter from \cite{hohpe2004enterprise}), or data size reduction (\eg Compress Content, Image Resizer from \cite{Ritter201736}).

Furthermore, there are 6 out of 153 patterns supporting the parallelization of the control flow (cf. OS-3).
The realization of parallel patterns and sequences, requires non-conditional forking (\eg Multicast \cite{Ritter201736}, Scatter-Gather, Wire Tap from \cite{hohpe2004enterprise}) and joining (\eg Join Router \cite{Ritter201736}) patterns.
In addition, heterogeneous load balancing can be realized structurally by a Load Balancer \cite{Ritter201736} and content-wise by an Aggregator pattern.

Since the pattern placement (cf. OS-4) denotes an extension of OS-2, all OS-2 patterns are applicable as well (\ie 16+12 out of 153).
The patterns that can be additionally pushed-down to the sender are categorized as early-quality of service (\eg Guaranteed Delivery \cite{hohpe2004enterprise}) and Early-Operation (\eg Adapter Flow, Content Sort, Custom Script, Find and Replace, Language Translator, Message Interceptor, Type Converter from \cite{Ritter201736}).
The patterns for push-back are according to their categories right-place (\eg Content Enricher \cite{hohpe2004enterprise}) and late-quality of service (\eg Idempotent Receiver, Resequencer from \cite{hohpe2004enterprise}, Commutative Endpoint \cite{Ritter201736}).

The reduction of interactions between endpoints (cf. OS-5), for which 13 out of 153 patterns can be used, can be categorized into flow control (\eg Back Pressure, Message Sampler, Message Throttler, Pause Operation, Timeout from \cite{Ritter201736}), resilience (\eg Circuit Breaker \cite{Ritter201736}), prevention (\eg Load Balancer), tolerance (\eg Delayed Redelivery, Failover Request Handler from \cite{Ritter201736}), filtering (\eg Request Collapsing, Request Caching, Request Partitioning from \cite{Ritter201736}, Aggregator).
Thereby most of the patterns from the flow control category reduce the number of requests per time, but not the general number of messages that lead to requests.

\subsection{Pattern Stratification}

The catalog of optimization strategies from \cref{sec:objectives} contains optimizations that have a similar effect, revert, support or inhibit each other.
Hence, we analzyed the dependencies between the optimizations and propose an order of their execution.
\begin{figure}[bt]
	\centering
	\includegraphics[width=0.9\columnwidth]{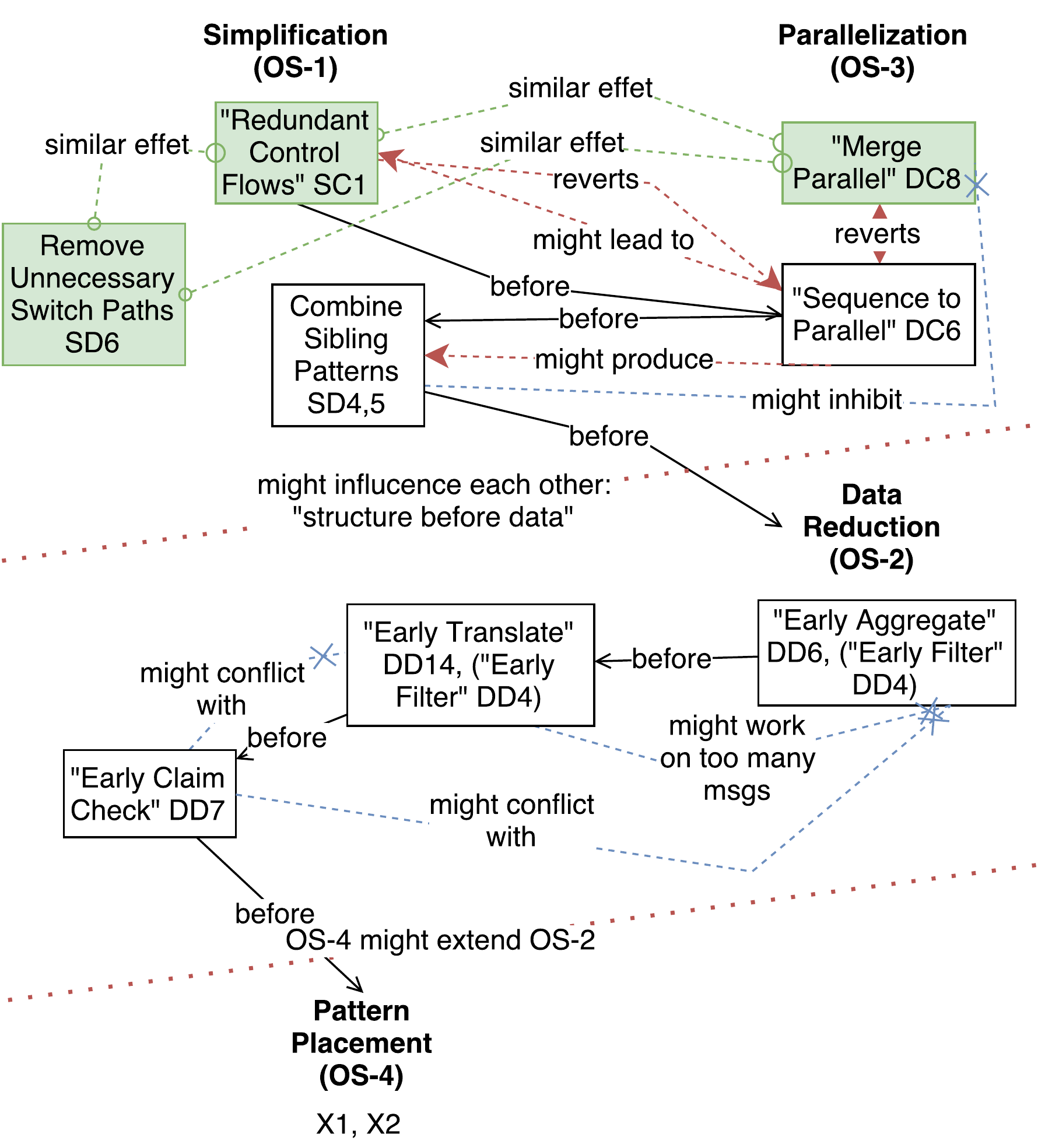}
	\caption{Optimization strategy application order.}
	\label{fig:osdependencies}
\end{figure}
\Cref{fig:osdependencies} shows the order, in which the optimization strategies can be executed.
The order is determined based on the conditions and impact of the optimization strategies.
According to that DC6 might lead to SC1, while SC1 might revert DC6, hence SC1 before DC6.
The dynamic control optimizations DC6 and DC8 revert each other, and thus require an order.
This order implicitly comes from the observation that  DC8, SC1, SD4,5 have a similar effect, and are thus related.
Since we know that SC1 comes before DC6, optimizations \{SC1, SD4,5, DC8\} have to be executed before DC6
DC6 should be executed before SD4,5, since it might produce SD4,5.
The static data optimization SD4,5 might inhibit DC8 in some cases, hence DC8 before SD4,5.
As an intermediate summary, we can formulate the general rule \enquote{structure before data}: strategies OS-1 and OS-3, before OS-2.

Now, the data reduction strategy optimizations DD6, DD4 have a similar effect, and thus can be applied in arbitrary order.
The same is valid for DD14, DD4.
Since DD14 might work on too many messages, DD6 should be before DD14 to reduce the number of messages.

The pattern placement strategy OS-4 might extend the reach of OS-2, hence OS-2 before OS-4.

The pattern placement strategy OS-4 might extend the reach of OS-2, hence OS-2 before OS-4.
The OS-5 strategies profit from all previous strategies and thus come last.

\section{Case Study: Italy Invoice Scenario (revisited)} \label{sec:case_study}

In this section, we study the application of the optimization strategies to the case of our motivating scenario.
Thereby the strategies are applied iteratively according to the sequence from \cref{fig:osdependencies}, unntil no further graph re-writing can be found.

\subsubsection{Simplify the Integration Process}
\begin{figure}[bt]
    \centering
    \includegraphics[width=1\linewidth]{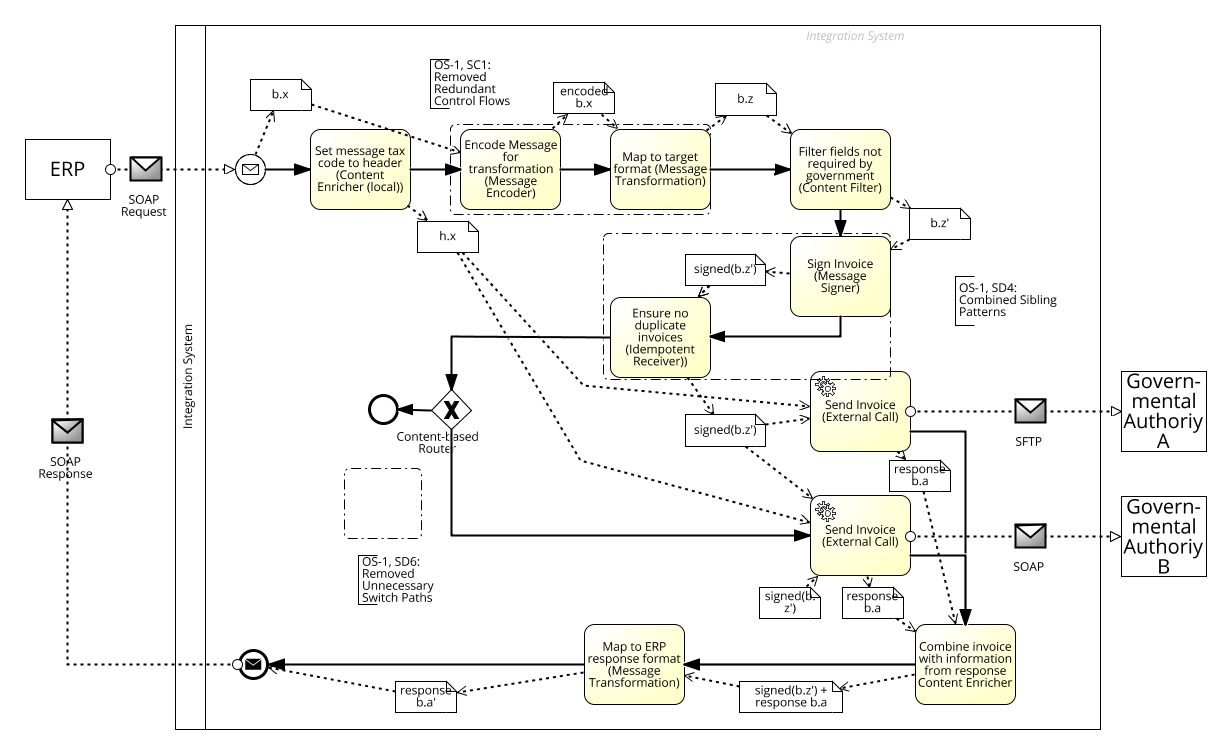}
    \caption{Integration process from \cref{fig:invoice_italy} after application of OS1 strategies.}
    \label{fig:invoice_simplified}
\end{figure}

The algorithm starts on the integration process \cref{fig:invoice_italy} by matching and applying the process simplification (OS-1) optimizations, whose sequence can be arbitrary.
For example, first, the redundant control flows (cf. SC1) starting from \emph{Encode Message for transformation} and ending with \emph{Map to target format} are identified and rewritten to a sequence without the multicast and join router patterns.
Then the dead paths (SC2) Content Enricher to End Event is removed.
And finally, the sibling patterns \emph{Sign Invoice} and \emph{Ensure no duplicate invoices} are identified and merged.
They are placed before the Content-based Router, thus pulled out of the conditional fork!
The resulting integration process is shown in \cref{fig:invoice_simplified}.
The original modeling complexity of $15$ patterns was reduced to $10$ patterns.


\subsubsection{Scale the Sub-Processes of the Integration Process}
Now the OS3 parallelzation strategies are applied on \cref{fig:invoice_simplified}.

- transformed sequence to parallel\\

\begin{figure}[bt]
    \centering
    \includegraphics[width=1\linewidth]{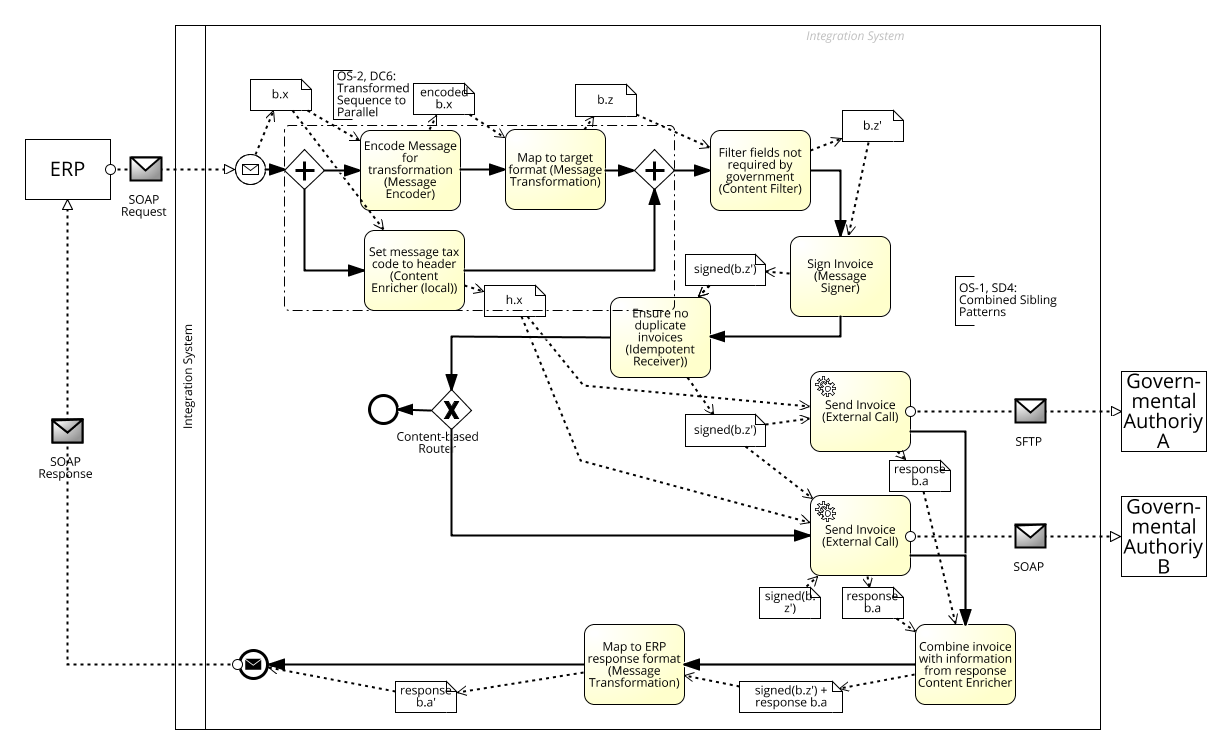}
    \caption{Integration process from \cref{fig:invoice_simplified} after application of OS3 strategies.}
    \label{fig:invoice_simplified_parallel}
\end{figure}


\begin{figure}[bt]
    \centering
    \includegraphics[width=1\linewidth]{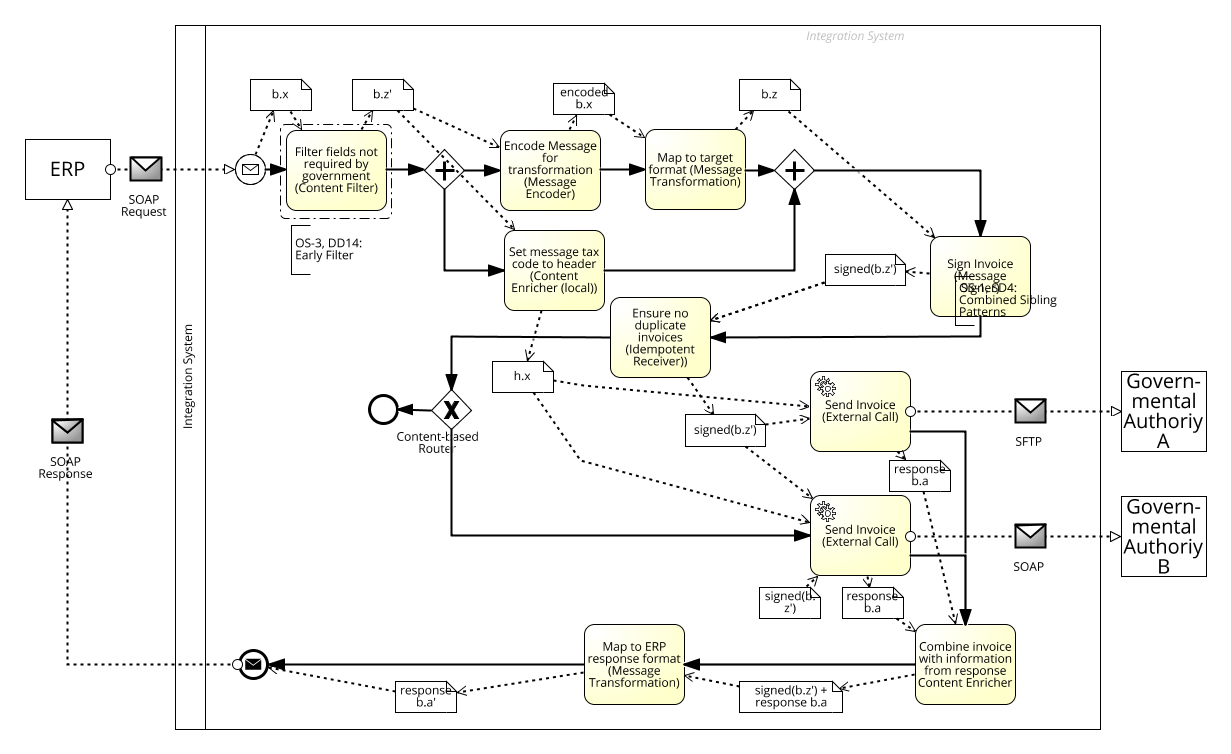}
    \caption{Integration process from \cref{fig:invoice_simplified_parallel} after application of OS2 strategies.}
    \label{fig:invoice_simplified_parallel_data}
\end{figure}

\subsubsection{Place Integration Sub-process to where they belong}
\begin{figure}[bt]
    \centering
    \includegraphics[width=1\linewidth]{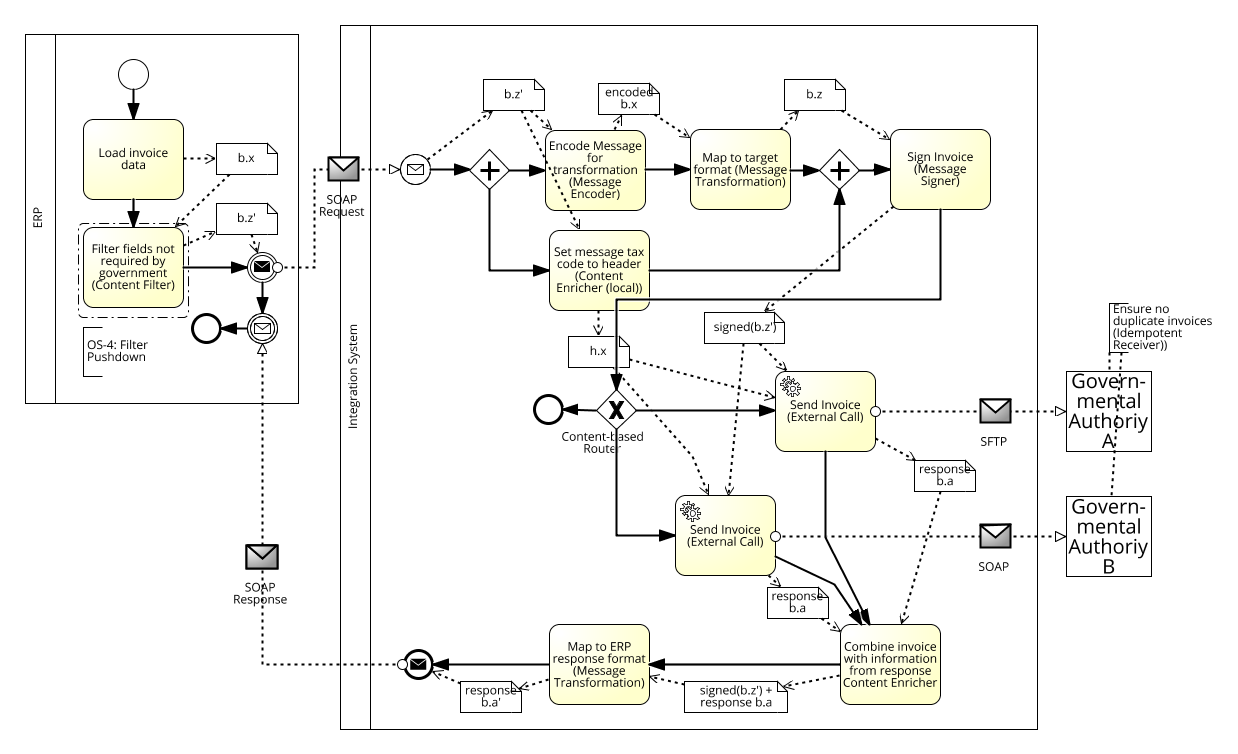}
    \caption{Integration process from \cref{fig:invoice_simplified_parallel_data} after application of OS4 strategies.}
    \label{fig:invoice_simplified_parallel_data_place}
\end{figure}

\Cref{fig:invoice_simplified_parallel_data_place} shows the original integration process from \cref{fig:invoice_italy} after the application of strategies OS-1 to 4 in the order denoted in \cref{fig:osdependencies}.
Notably, the Message Filter could be pushed down to the sender, which sends only relevant data for the receiver plus the data needed by the integration process.
The combination of Encode Message and Message Transformation can be executed in parallel to the Content Enricher.
Then the invoice is signed and sent to the authorities, which recognize duplicates and discard them.
Further OS-5 strategies could be applied, if necessary.


\section{Discussion} \label{sec:discussion}

In this work we collect and briefly explain a catalog of optimizations techniques from related domains that we transfered to EAI optimization strategies.
We formalize these optimizations and discuss their stratification as well as show the applicability for an extended scenario.

\bibliographystyle{abbrv}
\bibliography{../optimization}

\end{document}